\documentclass[fleqn,usenatbib]{mnras}
\usepackage[T1]{fontenc}
\usepackage{ragged2e,xcolor}
\usepackage{lmodern}
\DeclareRobustCommand{\VAN}[3]{#2}
\let\VANthebibliography\thebibliography
\def\thebibliography{\DeclareRobustCommand{\VAN}[3]{##3}\VANthebibliography}
\newcommand{\cii}{[C\,{\sc ii}]}

\newcommand{\hii}{H\,{\sc ii}}

\newcommand{\oiii}{[O\,{\sc iii}]}
\newcommand{\niii}{[N\,{\sc iii}]}

\newcommand{\oiiil}{[O\,{\sc iii}] 88\,$\mu{\rm m}$}
\newcommand{\ciil}{[C\,{\sc ii}] 158\,$\mu{\rm m}$}

\definecolor{referee}{RGB}{0,0,0}
\definecolor{referee2}{RGB}{0,0,0}

\usepackage{graphicx}	% Including figure files
\usepackage{amsmath}	% Advanced maths commands
\usepackage{amssymb}	% Extra maths symbols
\usepackage{newtxtext,newtxmath}
\graphicspath{./graphs/}
\title[Dust stack at $z > 8$]{Probing Infrared eXcess to Investigate Early-Universe Dust (PIXIEDust)}

\author[Bakx et al.]{
Tom J. L. C. Bakx$^{1}$\thanks{E-mail: tom.bakx@chalmers.se}, 
Hiddo S. B. Algera$^2$, 
Jean-Baptiste Jolly$^3$,
Clarke Esmerian$^{1}$,
Kirsten Knudsen$^{1}$,\newauthor{}
Laura Sommovigo$^{4}$,
Joris Witstok$^{5,6}$,
Stefano Carniani$^7$,
Jianhang Chen$^{8,9}$, 
Stephen Eales$^{10}$,\newauthor{}
Andrea Ferrara$^7$,
Yoshinobu Fudamoto$^{11,12,13}$,
Masato Hagimoto$^{14}$,
Takuya Hashimoto$^{15,16}$,\newauthor{}
Hanae Inami$^{17}$,
Akio K. Inoue$^{12,18}$,
Theo Khouri$^{1}$,
Ikki Mitsuhashi$^{19}$,
Gunnar Nyman$^{20}$,
Gustav Olander$^{1}$,\newauthor{}
Stephen Serjeant$^{21}$
Renske Smit$^{22}$,
Ilsang Yoon$^{23}$,
Jorge Zavala$^{24}$, 
Susanne Aalto$^1$, 
Caitlin M. Casey$^{25,5}$,\newauthor{}
Yoichi Tamura$^{15}$,
and
Wouter Vlemmings$^1$.
\\ 
$^{1}$Department of Space, Earth and Environment, Chalmers University of Technology, SE-412 96 Gothenburg, Sweden \\
$^2$Institute of Astronomy and Astrophysics, Academia Sinica, 11F of Astronomy-Mathematics Building, No.1, Sec. 4, Roosevelt Rd, Taipei 106319, Taiwan, R.O.C. \\
$^3$Max-Planck-Institut für extraterrestrische Physik, 85748 Garching, Germany \\
$^{4}$Center for Computational Astrophysics, Flatiron Institute, 162 5th Avenue, New York, NY 10010, USA\\
$^{5}$Cosmic Dawn Center (DAWN), Copenhagen, Denmark\\
$^{6}$Niels Bohr Institute, University of Copenhagen, Jagtvej 128, DK-2200, Copenhagen, Denmark\\
$^7$Scuola Normale Superiore, Piazza dei Cavalieri 7, I-56126 Pisa, Italy\\
$^8$European Southern Observatory (ESO), Karl-Schwarzschild-Strasse 2,Garching, Germany\\
$^9$Max-Planck-Institut für Extraterrestrische Physik (MPE), Giessenbachstraße 1, Garching, Germany\\
$^{10}$School of Physics and Astronomy, Cardiff University, The Parade, Cardiff, CF24 3AA, UK\\
$^{11}$Center for Frontier Science, Chiba University, 1-33 Yayoi-cho, Inage-ku, Chiba 263-8522, Japan\\ 
$^{12}$Waseda Research Institute for Science and Engineering, Faculty of Science and Engineering, Waseda University, 3-4-1 Okubo, Shinjuku, Tokyo 169-8555, Japan\\ 
$^{13}$National Astronomical Observatory of Japan, 2-21-1, Osawa, Mitaka, Tokyo, Japan\\
$^{14}$Department of Physics, Graduate School of Science, Nagoya University, Aichi 464-8602, Japan \\ 
$^{15}$Division of Physics, Faculty of Pure and Applied Sciences, University of Tsukuba, Tsukuba, Ibaraki 305-8571, Japan\\
$^{16}$Tomonaga Center for the History of the Universe (TCHoU), Faculty of Pure and Applied Sciences, University of Tsukuba, Tsukuba, Ibaraki 305-8571, Japan\\
$^{17}$Hiroshima Astrophysical Science Center, Hiroshima University, 1-3-1 Kagamiyama, Higashi-Hiroshima, Hiroshima 739-8526, Japan \\
$^{18}$Department of Physics, School of Advanced Science and Engineering, Faculty of Science and Engineering, Waseda University, 3-4-1 Okubo, Shinjuku, \\Tokyo 169-8555, Japan\\
$^{19}$Department for Astrophysical \& Planetary Science, University of Colorado, Boulder, CO 80309, USA\\
$^{20}$Department of Chemistry and Molecular Biology, University of Gothenburg, SE 413 90, Gothenburg, Sweden\\
$^{21}$School of Physical Sciences, The Open University, Milton Keynes, MK7 6AA, UK
$^{22}$Astrophysics Research Institute, Liverpool John Moores University, 146 Brownlow Hill, Liverpool L3 5RF, UK\\
$^{23}$National Radio Astronomy Observatory, 520 Edgemont Road, Charlottesville, VA 22903, USA\\
$^{24}$Department of Astronomy, University of Massachusetts, 619E LGRT, 710 N. Pleasant Street, Amherst, MA 01003, USA\\
$^{25}$Department of Physics, University of California, Santa Barbara, Santa Barbara, CA 93106, USA
}

\pubyear{2025}

\date{Accepted XXX. Received YYY; in original form ZZZ}

\begin{document}
\label{firstpage}
\pagerange{\pageref{firstpage}--\pageref{lastpage}}
\maketitle

% Abstract of the paper
\begin{abstract} 
Despite the implied presence of dust through reddened UV emission in high-redshift galaxies, no dust emission has been detected in the (sub)millimetre regime beyond $z > 8.3$. This study combines around two hundred hours of Atacama Large Millimeter/submillimeter Array (ALMA) and Northern Extended Millimeter Array (NOEMA) observations on ten $z > 8$ galaxies, revealing no significant dust emission down to a $1 \sigma$ depth of $2.0$, $2.0$, and $1.5 \,\mu$Jy at rest-frame 158, 88~$\mu$m, and across all the data, respectively. This constrains average dust masses to be below $< 10^{5}$~M$_{\odot}$ at $3 \sigma$ and dust-to-stellar mass ratios to be below $3.7 \times{} 10^{-4}$ (assuming $T_{\rm dust} = 50$~K and $\beta_{\rm dust} = 2.0$). 
Binning by redshift ($8 < z < 9.5$ and $9.5 < z < 15$), UV-continuum slope ($\beta_{\rm UV} \lessgtr -2$) and stellar mass ($\log_{10} M_{\ast}/{\rm M_{\odot}} \lessgtr 9$) yields similarly stringent constraints. Combined with other studies, these results {\color{referee} are consistent with} inefficient dust build-up in the $z > 8$ Universe, likely due to inefficient supernova production, limited interstellar grain growth and/or ejection by outflows. We provide data and tools online to facilitate community-wide high-redshift dust searches.
\end{abstract}

% Select between one and six entries from the list of approved keywords.
% Don't make up new ones.
\begin{keywords}
dust, extinction – galaxies: evolution – galaxies: formation – galaxies: high-redshift – submillimetre: galaxies.
\end{keywords}

%%%%%%%%%%%%%%%%%%%%%%%%%%%%%%%%%%%%%%%%%%%%%%%%%%

%%%%%%%%%%%%%%%%% BODY OF PAPER %%%%%%%%%%%%%%%%%%

\section{Introduction}
\begin{table*}
    \centering
    \caption{Properties of $z > 8$ galaxies}
    \label{tab:sourceProperties}
    \begin{tabular}{lcccccccc}
    \hline
Source & RA & Dec & Redshift & $M_{\rm UV}$ & $\beta_{\rm UV}$ & $\log_{10}$ $M_{\ast}$ & Size & $\mu$ \\
& [hms] & [dms] & & & & [$M_{\odot}$] &  [pc] \\
\hline
REBELS-24 & 10:00:31.89 & $+$01:57:50.20 & 8.21 & $-$22.0 & $-$1.5$^{+0.5}_{-1.1}$ & 8.89 $\pm$ 0.68 & - & -\\
GS-z9-3 & 03:32:34.99 & $-$27:49:21.60 & 8.228 & $-$19.8 & - & 9.19 $\pm$ 0.07 & - & - \\
ID4590 & 07:23:26.26 & $-$73:26:57.04 & $8.496$ & $-$19.6 & $-1.70 \pm 0.07$ & ${8.82 \pm 0.31}$ &  $280 \pm 50$ & 8.69 \\
MACS1149-JD1 & 11:49:33.58 & $+$22:24:45.70 & 9.110 & $-$19.6 & $-2.20 \pm 0.01$ & $8.20 \pm 0.05$  & $332 \pm 54$ & 10.5 \\ 
GS-z9-0 & 03:32:26.94 & $-$27:46:28.72 & 9.433 & $-20.4$ & $-2.54 \pm 0.02$ & $8.17 \pm 0.05$ & $110 \pm 9$ & - \\
GHZ-1 & 00:14:02.86 &  $-$30:22:18.69 & 9.875 & $-$20.0 & $-1.79 \pm 0.05$ & $9.20 \pm 0.30$ & $500 \pm 20$ & - \\
% COSMOS-20646 & 10:00:19.64 & $+$02:15:45.90 & $9.77 \pm 0.19$ & $-22.1$ & $-0.62 \pm 0.12$ & $10.9 \pm 0.2$ & - \\
% 2140+0241-37 & 21:39:32.58 & $+$02:41:06.51 & $\sim10.5$ & $-22.2$ &  &  &  & -  \\ 
GN-z11 & 12:36:25.46 & $+$62:14:31.40 & 10.603 & $-21.5$ & $-$2.36 $\pm$ 0.10 & 8.73 $\pm$ 0.06  & $64 \pm 20$ & - \\
GS-z11-0 & 03:32:39.54 & $-$27:46:28.67 & 11.58 & $-19.3$ & $-2.18 \pm 0.09$ & $8.67 \pm 0.13$ & $80 \pm 20$ & - \\
GHZ-2 & 00:13:59.76 & $-$30:19:29.22 & 12.333 & $-$20.5 & $-2.46 \pm 0.08$ &  $9.05 \pm 0.18 $  & $105 \pm 9$ & 1.3 \\
GS-z14-0 & 03:32:19.90 & $-$27:51:20.27 & 14.178 & $-$20.8 & $-2.20 \pm 0.07$ & $8.70 \pm 0.50$  & $260 \pm 20$ & $< 1.2$ \\ 
\hline
\multicolumn{9}{c}{{\bf Not stacked}} \\
\hline
MACS0416-Y1 & 04:16:09.40 & $-$24:05:35.47 & 8.312 & $-$20.8 & $-1.72 \pm 0.50$ & $8.60 \pm 0.10$ & $390 \pm 50$ & 1.5 \\ 
SPT-0615-JD & 06:15:55.10 & $-$57:46:20.19 & 9.625 & $-$17.8  & $-2.70 \pm 0.10$ & $7.47\pm 0.18$ & - & 120 \\ 
COS-z-0 / COS-z12-1 & 09:58:55.23 & $+$02:07:16.82 & $12.25 \pm 0.25$ & $-$22.2 & $-$1.78 $\pm$ 0.23 & $9.60 \pm 0.11$ & $420 \pm 70$ & - \\ 
\hline
\end{tabular} 
\raggedright \justify \vspace{-0.2cm}
\textbf{Notes:} 
Col. 1: Source name. 
Col. 2: Right Ascension in [hms] units.
Col. 3: Declination in [dms] units.
Col. 4: Redshift, if errors are listed, the redshift is derived from photometric observations. Note that the redshifts of REBELS-24 and GS-z11-0 are based on the modest resolution of JWST spectroscopy, and are thus accurate to two digits.
Col. 5: M$_{\rm UV}$ corrected for lensing.
Col. 6: The slope of the UV continuum. For GS-z9-3, the $\beta_{\rm UV}$ is not available, but the DAWN JWST Archive (DJA) and its subsequent analysis provide physical estimates of $A_V = 0.29$ and $L_{\rm IR} = 6.9 \times{}10^{10}$~$L_{\odot}$. 
Col. 7: The stellar mass estimate, corrected for lensing. The references can be found in the Appendix Section~\ref{sec:appStackedSources}.
Col. 8: The rest-frame UV / optical size of the source corrected for lensing.
Col. 9: Lensing magnification.
\end{table*}

Despite containing only 0.1 to 0.5 per cent of all baryons, dust plays a disproportionately important role in our Universe and the galaxies within it \citep{Peroux2020}. The bulk of gas available for star formation (i.e., molecular gas) forms on the chemically active surfaces of dust grains, and its thermal radiation provides an efficient cooling method to facilitate the gravitational collapse of gas into stars. Observationally, dust is seen through the reddening of stellar emission at optical wavelengths, as well as the subsequent emission from these heated dust grains through radiation in (sub-)mm wavelengths \citep[e.g.,][]{daCunha2015}. As dust studies extend to higher redshifts, %At a redshift $z = 8$, 
the time since the Big Bang starts to approach the typical timescales of the dust production mechanisms themselves. 
Stars with $\sim 8$~M$_{\odot}$ require at least 30 to 100 million years to reach the Asymptotic Giant Branch phase (AGB; \citealt{DwekCherchneff2011,Lugaro2012,Boyer2025}), and while supernovae can occur on faster timescales, the destructive effect of the reverse shock on the total dust yield is expected to only leave between 10 to 30~per cent intact \citep{Schneider2024Review}. The surprising discovery of dust at $z > 7$ \citep{watson2015dusty,knudsen2017merger} has created optimism towards a direct detection of dust emission in the high-$z$ Universe, and the subsequent several hundred hours of interferometric observations now provide a unique opportunity to test the dust formation mechanisms as well as the effect of dust on galaxy evolution.
% had 600 million years to form the observed dust. Meanwhile, 
% dust formation pathways have similar timescales as the age of the Universe at that time . 

Even the most distant galaxy candidates show indications of dust through their reddened UV-continuum emission out to redshifts $z \sim 14$ \citep[e.g.,][]{Heintz2023,Carniani2024highzSources,Kokorev2025}. While spectroscopic confirmation of these sources is key to rule out low-redshift interlopers \citep{Fujimoto2023,Zavala2023Masquerading}, the first three years of \textit{James Webb Space Telescope} (\textit{JWST}) operation have already confirmed many of these systems as true high-$z$ galaxies \citep[e.g.,][]{Harikane2024}. Beyond the direct reddening of the UV-continuum slope (i.e., the $\beta_{\rm UV}$), the rest-frame UV emission of galaxies can also provide further insight into dust composition, including a characteristic 2175~\AA{} absorption feature indicating the presence of carbonaceous dust specifically \citep{Witstok2023,Markov2023}. Using rest-frame UV emission to characterize dust properties requires resolved observations combined with radiative transfer models \citep{casey2014}, which are beyond even the capabilities of the \textit{JWST}. Model dependencies significantly affect results \citep{Martis2019,Markov2023}, and rest-frame UV observations are sensitive to the assumed stellar population age and metallicity \citep{Barisic2017}. %The abundance of unexpectedly massive and evolved galaxies at redshifts beyond $z = 8$ \cite[e.g.,][]{Boylan-Kolchin2022,Weibel2025} furthermore calls into question galaxy formation pathways. % with one potential reason for their high UV-based stellar masses requiring dust-driven outflows \citep{Carniani2024EventfulLife,Ferrara2024EventfulLife}. 

Despite the clear evidence for dust in the $z> 8$ Universe through the UV-continuum slope \citep[e.g.,][]{Casey2024} and the suggested presence of carbonaceous dust grains at $z > 7$, direct detections of dust emission from individual galaxies at these extreme redshifts remain scarce \citep{Tokuoka2022,Bakx2023,Yoon2023,Schouws2024OIII,Carniani2024highzSources,Mitsuhashi2025}. Only one source, MACS0416\_Y1, has evidence of dust emission seen in the sub-mm regime at $z = 8.3$ \citep{Tamura2019,Bakx2020CII,Harshan2024}. 
Since the identification of $z > 8$ objects with the \textit{Hubble Space Telescope} (\textit{HST}), the Atacama Large Millimetre/submillimetre Array (ALMA) has been used for several tens to hundreds of hours to characterize the dust emission of distant galaxies \citep[e.g.,][]{Tamura2019,Tamura2023,Bakx2020CII,Bakx2023,Inami2022,Fujimoto2024,Fudamoto2024}. With dust readily detected at $z = 6 - 8$ \citep{Riechers2013,watson2015dusty,LeFevre2020,Bethermin2020,Inami2022}, the scarcity in sub-mm detections becomes an intriguing question with cosmological implications \citep{Lesniewska2019}. {\color{referee} Building upon the advances of composite images of galaxies at cosmic noon ($z = 1 - 5$; \citealt{Bouwens2016,Bouwens2020,Dunlop2017,McLure2018,jolly2025})}, innovative stacking analyses of far-infrared data have revealed elevated infrared luminosity density extending to $z \sim 8$ to 10 \citep{Viero2022,Algera2023,Ciesla2024}, and can provide an additional avenue to test the comprehensive picture of dust properties at high redshifts. 
{\color{referee2} At higher redshifts, even ground-based observations are able to probe closer to the peak of the dusty spectrum, which are expected to be warmer than in the $z = 0$ Universe \citep{Schaerer2015,Faisst2017,Liang2019,Bakx2020CII,Sommovigo2021,Sommovigo2020hotdustorigins}. The (sub-)mm K-correction furthermore increases the sensitivity of observations by roughly 1.5 to five-fold, depending on the dust temperature \citep{blain1999,blain2002,hodge2020}.}
However, stacking exercises are sensitive to interlopers, and it is important to have high confidence in the true high-redshift nature of distant galaxies \citep[c.f.,][]{Zavala2023Masquerading}. 

In this study, we examine the constraints on dust physics provided by existing deep ($t_{\rm int} > 1$~hr) (sub)mm observations of galaxies with robust redshifts in the $z > 8$ Universe. Through a comprehensive framework, publicly available at \url{http://github.com/tjlcbakx/high-z-dust-stack}, a stacking exercise in flux, dust mass, and dust-to-stellar mass ratio explores our best view yet on obscured star-formation in the $z > 8$ Universe through both rest-frame UV and optical observations, and those at (sub)mm wavelengths. The data consists of a combination of previously-published results and unpublished results for ten galaxies. Throughout this paper, we assume a flat $\Lambda$-CDM cosmology with the best-fit parameters derived from the \textit{Planck} results \citep{Planck2020}, which are $\Omega_\mathrm{m} = 0.315$, $\Omega_\mathrm{\Lambda} = 0.685$ and $h = 0.674$.

\section{Sample and Observations}
This study focuses on a sample of galaxies (Table~\ref{tab:sourceProperties}) beyond $z > 8$ with substantial observing time ($\gtrapprox 1$~hour) with a sensitive interferometer, in particular Northern Extended Millimeter Array (NOEMA) and ALMA. These objects are selected from deep rest-frame UV and optical observations using the \textit{HST} and \textit{JWST}, which search for the characteristic Lyman-break feature and/or emission lines that reveal the high-redshift nature of these objects. 

These distant objects are often found at the edge of the detection limits of \textit{HST} and \textit{JWST}, which increase the risk of including interlopers such as local brown dwarfs, $z \approx 1$ quiescent galaxies \citep{Harikane2025}, and $z \approx 4$ dusty Balmer-break galaxies \citep{Naidu2022b,Zavala2023Masquerading}. To avoid the confusing effect of interlopers in this stacking experiment, this study is limited to all objects that are robustly confirmed to lie at $z > 8$. For completeness, we include (but do not stack) the singular source at $z > 8$ that to date has been detected in dust, MACS0416\_Y1 \citep{Tamura2019}. %This source is not added to the stack, as the scope of this paper is to investigate the nature of sub-mm undetected $z > 8$ galaxies.

In the Appendix~\ref{sec:appStackedSources}, we discuss the notable properties of each object in order of increasing redshift, including the existing rest-frame UV, optical, and far-infrared observations. All data except those for GN-z11, which was taken with NOEMA, have been re-calibrated and re-imaged for consistency. These are calibrated using the provided scripts with the suggested \textsc{CASA} pipeline version. The subsequent data reduction steps use \textsc{TCLEAN} \citep{TheCASATeam_2022} with a Hogbom deconvolver and a natural (\textit{robust} = 2) weighting. In the case when the archival data are not yet published previously, we provide additional details on the observational set-up in the same Appendix. 

\begin{figure}
    \centering
    \includegraphics[width=\linewidth]{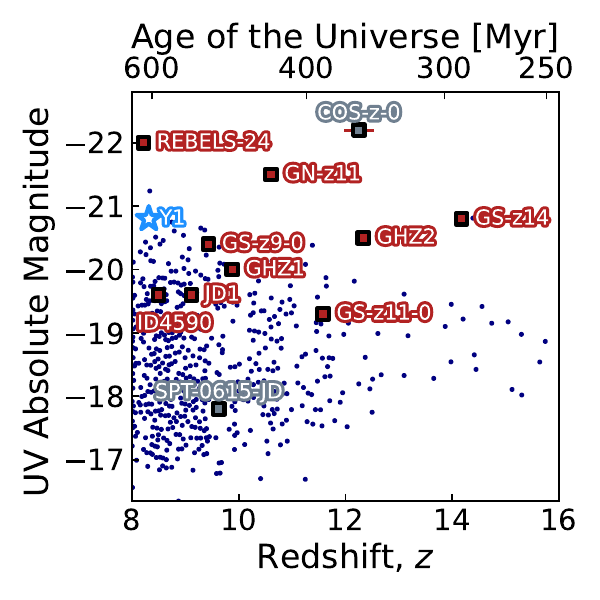}
    \caption{UV Magnitude of our stacking sample of spectroscopically confirmed z>8 galaxies (\textit{red squares}) as a function of their redshift and age of the Universe. Galaxies with deep observations that are not stacked are shown in blue (Y1; \citealt{Tamura2019}) and grey squares (COS-z-0 and SPT-0615-JD), and compared to galaxy candidates with photometric redshift estimates from the GOODS-N and GOODS-S fields shown in \textit{blue points} \citep{Hainline2024}.}
    \label{fig:MuvVsZ}
\end{figure}

The source details are shown in Table~\ref{tab:sourceProperties}, with their sub-mm observations detailed in Table~\ref{tab:observationProperties}. 
The sources that will be stacked are typically observed for an on-source time between 1.4 and 80.6 hours for a total of $\sim175$~hrs of on-source time, with the total necessary observing time roughly 30 to 50~\% higher due to calibration and overheads. The beam sizes of the observations are larger than the reported physical sizes of the sources, ranging between 0\farcs{}37 and 1\farcs{}35. 
Figure~\ref{fig:MuvVsZ} shows the UV absolute magnitude of these sources relative to other UV-detected $z > 8$ galaxies. Individual poststamps of the (sub)mm emission of each of the sources can be seen in Figure~\ref{fig:continuumImages}.

\begin{table*}
    \centering
    \caption{Description of the available ALMA and NOEMA data}
    \label{tab:observationProperties}
    \begin{tabular}{lccccccccc}
    \hline
Source & $f_{\rm obs}$  & $S_{\nu}$ & t$_{\rm obs}$ & Beam size & Proposal ID \\
 & [GHz] & [$\mu$Jy] & [hr] & [$" \times "$]  \\ \hline

REBELS-24  & 202  & $-3.3 \pm 9.3$ & 2.2 & 1.26 $\times$ 1.06 & 2018.1.00236.S, 2019.1.01634.L \\
GS-z9-3 & 349 & $12.2 \pm 16.1$ & 2.2 & 1.00 $\times$ 0.74 & 2017.1.00486.S \\
ID4590 & 200 & $-7.6 \pm 11.8$ & 1.7 & 1.35 $\times$ 1.25 & 2021.A.00022.S \\
& 360 & $-17.8 \pm  22.0$ & 1.4 & 0.73 $\times$ 0.59 & 2021.A.00022.S \\
MACS1149-JD1 &  188 & $4.1 \pm 5.8$ & 4.6 & 0.77 $\times$ 0.66 & 2017.A.00026.S \\
& 335 & $13.9 \pm 8.1$ & 14.2 & 0.37 $\times$ 0.31 & 2015.1.00428.S, 2018.1.00616.S \\
GS-z9-0 & 307 & $-4.5 \pm 6.9$ & 4.6 & 0.90 $\times$ 0.64 & 2022.1.01401.S \\
GHZ-1 & 248 & $12.2 \pm 5.7$ & 7.0 & 0.80 $\times$ 0.60 & 2021.A.00023.S \\
% COSMOS-20646 & 320 & $<$26.0 & 5.7 & 0.41 $\times$ 0.38 & 2019.1.00397.S, 2022.1.01562.S \\
% 2140+0241-37 & 308 & $<$35.6 & 1.9 & 0.41 $\times$ 0.38 & 2019.1.00397.S, 2022.1.01562.S \\
GN-z11 &  161 & $4.0 \pm 12.9$ & 80.6 & 0.41 $\times$ 0.38 & XACE, XBCE, XCCE, XDCE, S16CQ, W17FD, W18FD \\
GS-z11-0 & 281 & $7.1 \pm 4.2$ & 19.2 & 0.88 $\times$ 0.69 & 2023.1.00336.S \\
GHZ-2 & 248 & $1.5 \pm 3.4$ & 12.2 & 0.51 $\times$ 0.43 & 2021.A.00020.S, 2023.A.00017.S \\
& 413 & $15.8 \pm 16.6$ & 8.2 & 0.71 $\times$ 0.62 & 2023.A.00017.S, 2024.1.01645.S, 2024.1.01771.S \\
GS-z14-0 &  125 & $-2.3 \pm 3.0$ & 5.6 & 1.27 $\times$ 0.93 & 2024.A.00007.S \\
& 224 & $0.4 \pm 5.0$ & 8.2 & 0.82 $\times$ 0.66 & 2023.A.00037.S \\
\hline
\multicolumn{6}{c}{{\bf Not stacked}} \\
\hline
MACS0416-Y1$^{\dagger}$ & 204 & $12 \pm 6$ & 6.5 & 0.44 $\times$ 0.31 & 2017.1.00225.S, 2019.1.01350.S \\
& 364 & $137 \pm 26$  & 21.5 & 0.14 $\times$ 0.11 & 2016.1.00117.S, 2017.1.00225.S, 2017.1.00486.S, 2018.1.01241.S \\
& 465 & $207 \pm 65$  & 1.6 & 0.84 $\times$ 0.63 & 2019.1.00343.S \\
SPT-0615-JD$^{\dagger}$ & 307 & $<$15.2 & 10.0 & 0.48 $\times$ 0.45 & 2018.1.00295.S, 2019.1.00327.S \\
COS-z-0$^{\dagger}$ / COS-z12-1 & 258 & $<$15.4 & 5.4 & 0.51 $\times$ 0.42 & 2023.A.00003.S \\
\hline
\end{tabular}
\raggedright \justify \vspace{-0.2cm}
\textbf{Notes:} 
Col. 1: Source name. 
Col. 2: The frequency of the sub-mm observations.
Col. 3: The observed flux density at the source position, given as the observed flux and error for all sources but SPT-0615-JD and COS-z-0/COS-z12-1, where the flux is likely too extended and confused by a nearby bright galaxy, respectively.
Col. 4: The total on-source time of the observations as listed on the ALMA Science Archive, not including overheads. For NOEMA data, the on-source time is taken from \citet{Fudamoto2024}.
Col. 5: ALMA / NOEMA Proposal IDs.
$^{\dagger}$ These sources are not included in the stacking analysis.
\end{table*}

%  & 02:17:01.35 & -05:09:59.78 & $9.54 \pm 0.10$ & $-22.1$ & $1.41 \pm 0.06$ & $11.0 \pm 0.3$ & \\ % Tacchella+ 

% Seiji project: 2022.1.01562.S \\
% 0925+1360 \\
% 0242+0716 \\
% 0853+0310 \\
% 1433-1804 \\
% 1525-1717 \\
% 1525+0960 \\
% 2008-6610 \\
% 2203+1851 & & & 7.838\\
% C2020\_852845 \\
% COS20646 \\
% COS47074 \\
% PAR335+13-251 \\
% %PAR0456-2203-473 & & & low-z\\
% PAR0750+2917-1736 \\
% PAR0756+3043-92 \\
% PAR0956-0450-684 \\
% PAR0956+2847-169 & & & 8.230 / 8.205\\
% PAR0956+2847-1130 & & & 8.490 \\
% PAR2139+0241-1709 & & & low-z \\
% PAR2346-0021-164 & & & low-z \\
% UDS18697 & & & low-z \\
% GS-z8.22 & & & 8.22 \\ % Project by Sun, Fengwu https://almascience.eso.org/aq/?project_code=2023.1.00391.S\&projectCode=2023.1.00391.S
% RXJ2129-z951 & & & 9.51 \\ % Project by Yoshinobu Fudamoto https://almascience.eso.org/aq/?project_code=2023.1.00235.S&projectCode=2023.1.00235.S

\section{Stacking analysis}
Stacking is a method that can increase the fidelity of existing observations. The underlying assumption is that the composite images are representative of the same population \citep{spilker2014,Hagimoto2023,Reuter2023}. The galaxies in this sample span a large range in redshift ($z = 8 - 14.5$), stellar mass ($\log_{10} M_{\ast} / M_{\odot} =$~7.4 to 9.2), and a $\beta_{\rm UV}$ between $-2.7$ and $-1.5$. Each of these parameters are likely correlated with the total amount of dust present in these galaxies. In an effort to test the effects of this potential variation, this study attempts different weighting schemes in the stacking across the complete sample, as well as in redshift, stellar mass, and UV-continuum slope bins. 

Throughout this study, we use the \textsc{LineStacker} tool to perform the stacking \citep{Jolly2020}. 
Given the large variation in observations from different observation cycles and necessary weighting schemes, the stacking is performed in the image plane \citep{Lindroos2015}. This furthermore facilitates the combination of NOEMA and ALMA data, as well as making the tool readily available for the public. 
To this end, all images are produced with the same pixel size. A smoothing to a common beam larger than the largest individual beam was attempted, but resulted in strong ($\sigma_{\rm smoothed} > 3 \times{} \sigma_{\rm unsmoothed}$) increases in the per-beam noise of the sources with the highest-resolution observations. Since the sources are expected to be unresolved on the scale of even the smallest beams, which still extend over more than a kiloparsec at $z = 9$ \citep{Morishita2024SubkpcScales}, we focus solely on the pixels centered on the source position to achieve the highest fidelity stack. The flux-based stacks are not corrected for magnification, while the dust mass stacks directly measure a physical quantity, and thus are corrected for magnification. The stacks of the dust-to-stellar mass ratio assume the same magnification factor for both the stellar as the dust mass, and thus are ``unaffected'' by the magnification correction. The individually-estimated dust masses, derived as discussed below, are given in Table~\ref{tab:Dustlimits}, and the results of the complete stacking study are given in Table~\ref{tab:stacking}.

\begin{table}
    \centering
    \caption{Dust limits of individual sources assuming $T_{\rm dust} = 50$~K}
    \label{tab:Dustlimits}
    \begin{tabular}{llll}
    \hline
Source &{$M_{\rm dust}$}  & {$M_{\rm dust}$/$M_{\ast}$} & $M_{\rm dust, opt}$ \\
 & {[$10^6$~M$_{\odot}$]} & $\times 10^{-3}$ &  [$10^6$~$M_{\odot}$] \\ \hline
REBELS-24    & $<$7.9 & $<$10 & \textit{1.38} \\
GS-z9-3    & $<$2.8 & $<$1.8 &  \textit{0.18} \\
ID4590  & $<$0.39 & $<$6.7 & 1.03 \\
MACS1149-JD1 & $<$0.09 & $<$0.60 & 0.50 \\
GS-z9-0 & $<$1.55 & $<$11 & 0.13 \\
GHZ-1 & $<$2.13 & $<$1.3 & 25.68 \\
% COSMOS-20646  & $<$1.77 & $<$0.000022 & 2.46 \\
SPT-0615-JD & $<$0.0094 & $<$0.32 & - \\
% 2140+0241-37 & $<$2.62 & - & - \\
GN-z11 & $<$19.2 & $<$36 & 0.14 \\
GS-z11-0 & $<$1.13 & $<$2.4 & 0.35 \\
COS-z-0 / COS-z12-1  & $<$1.68 & $<$0.42 & 18.33 \\
GHZ-2 & $<$0.94 & $<$0.83 & 0.18 \\
GS-z14-0 & $<$2.42 & $<$4.7 & 3.55 \\
\hline
\end{tabular}
\raggedright \justify \vspace{-0.2cm}
\textbf{Notes:} 
Col. 1: Source name. 
Col. 2: The $3 \sigma$ dust mass limit through the combined infrared fits, corrected for gravitational lensing, assuming a 50~K dust temperature and a $\beta_{\rm dust} = 2$.
Col. 3: The $3 \sigma$ dust-to-stellar mass ratio, where stellar masses are publically available (see Table~\ref{tab:sourceProperties} and the individual discussion of sources in the Appendix Section~\ref{sec:appStackedSources}.
Col. 4: The optically-derived dust masses predicted from $\beta_{\rm UV}$ and optical sizes. The mass for REBELS-24 and GS-z9-3 are derived with an ad-hoc 100~pc size based on typical REBELS-sizes from \citet{Ferrara2022REBELS}, and $\tau_{1500} = 0.29/1.086$ for GS-z9-3, while for SPT-0615-JD no representative radius is available.
\end{table}

\begin{table*} \color{referee2}
    \centering
    \caption{Stacking results for the complete sample and in separate bins}
    \label{tab:stacking}
    \begin{tabular}{lllllll}
    \hline
 & & $S_{\nu}$ & $M_{\rm dust}$ & $M_{\rm dust}$/$M_{\ast}$ & $M_{\rm dust, opt}$ & $M_{\rm dust, flux}$ \\ 
&& [$\mu$Jy] & [$10^4$~M$_{\odot}$] & $\times 10^{-4}$ & [$10^4$~M$_{\odot}$] & [$10^4$~M$_{\odot}$] \\
 \hline
r.f. 88~$\mu$m & ($8$) & $< 6.1$ & $< 31.0$ & $< 6.7$ & - &  61 \\%\textit{125} \\
r.f. 158~$\mu$m & ($5$) & $< 6.1$ & $< 8.6$ & $< 5.5$ & - &  61 \\%\textit{621} \\
All & ($14$) & $< 4.6$ 	                              & $< 9.1 $ & $< 3.7$   & 57  & 46 \\ \hline
$z > 9.5$ & ($7$) & $< 5.3$ 	                          & $< 66.1$ & $< 6.5$   & 292  & 53   \\
$z < 9.5$ & ($7$)& $< 10.0$ 	                      & $< 9.1 $ & $< 4.6$   & 52  & 99  \\
$M_{\ast}  > 10^9 {\rm\, M_{\odot}}$ & ($4$) & $< 9.2$   & $< 88.5$ & $< 7.3$ & 433   & 91  \\
$M_{\ast} <  10^9 {\rm\, M_{\odot}}$ & ($10$)& $< 5.5$   & $< 9.1$  & $< 4.5$ & 53 & 55  \\
$\beta_{\rm UV} > -2$ & ($6$) & $< 9.8$ 	              & $< 42.7$ & $< 5.9$   & 50  & 97  \\
$\beta_{\rm UV} < -2$ & ($8$)& $< 5.4$ 	              & $< 9.1$  & $< 4.8$   & 183 & 54   \\ 
\hline
\end{tabular}
\raggedright \justify \vspace{-0.2cm}
\textbf{Notes:} 
Col. 1: Stacking classification followed by the number of stacked sources in brackets.
Col. 2: The $3 \sigma$ flux density limit through the combined infrared fits.
Col. 3: The $3 \sigma$ dust mass limit through the combined infrared fits assuming a 50~K dust temperature and a $\beta_{\rm dust} = 2$.
Col. 4: The $3 \sigma$ dust-to-stellar mass ratio, where stellar masses are publicly available as mentioned in Table~\ref{tab:sourceProperties}.
Col. 5: The weighted-sum of the optically-derived dust masses predicted from $\beta_{\rm UV}$ and optical sizes. The optical dust mass estimates for the rest-frame 88 and 158~$\mu$m are not listed, as it double-counts sources that both have 88 and 158~$\mu$m observations.
Col. 6: The flux density-based dust masses estimated from the average observing frequency and redshift of our sources, which provides a sample-based estimate on the dust mass that is less sensitive to individual outlier sources. 
\end{table*}

\subsection{Flux-weighted stacking}
\label{sec:sensitivityweightedstacks}
The most common approach at stacking to detect dust emission is through the combination of observations at the same rest-frame wavelength. Although the observed frequencies might differ, these should still probe a similar part of the dusty spectrum, and can be combined directly. The stack is created using the equation
\begin{equation}
S_{\nu} = \frac{\sum_i S_{\nu, i} \times{} \sigma_i^{-2}}{\sum_i \sigma_i^{-2}} \label{eq:snrstack}
\end{equation}
for each pixel. Here, $S_{\nu,i}$ is the flux density at rest-frame frequency $\nu$ for observation $i$, and $\sigma_i$ the estimated noise around the position of the source. The $\sigma_i$ for each band is given as the uncertainty in the flux estimates in Table~\ref{tab:observationProperties}. 

\begin{figure*}
\centering
    \includegraphics[width=0.33\linewidth]{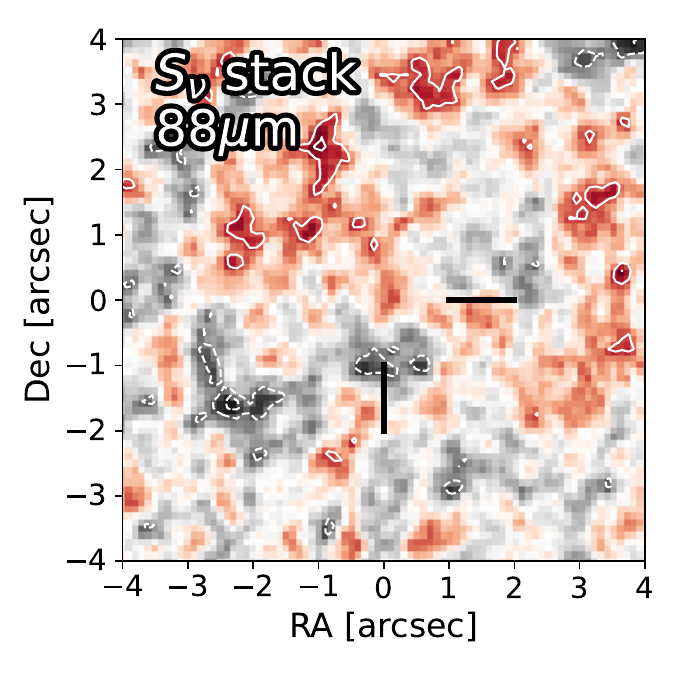}
    \includegraphics[width=0.33\linewidth]{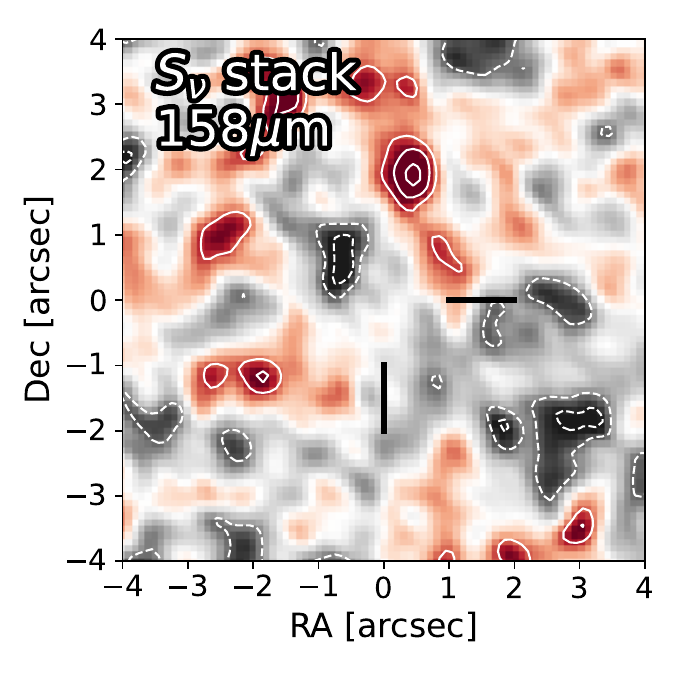}
    \includegraphics[width=0.33\linewidth]{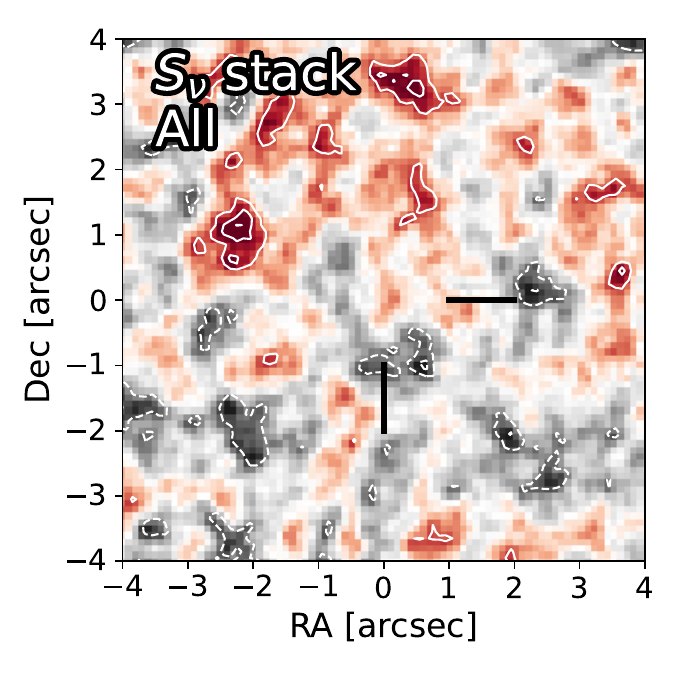}
    \caption{ The left-hand side and middle panels show the stacks of the rest-frame 88 and 158~$\mu$m emissions of eight and five observations, respectively. The right-hand side panel shows the combined stacked emission across all fourteen observations. The continuum weighting is based on the standard deviation of the continuum maps, with dashed and solid contours indicating 2, 3, 4... $\sigma$ in 8 by 8 arcsecond poststamps. The $3 \sigma$ depths of the maps are detailed in Table~\ref{tab:stacking}. The black indicators guide the eye to the central region where the rest-frame UV / optical emission of the sources is detected.}
    \label{fig:88_158um}
\end{figure*}

Figure~\ref{fig:88_158um} shows the stacked emission from the individual sources at 88 and 158~$\mu$m rest-frame emissions, providing deep upper limits on the total dust emission seen in the data. For completeness, a stack of the dust continuum observations across all frequencies is also shown in the right-hand side panel, where fluxes are directly stacked without any subsequent conversion to a common rest-frame wavelength. 

No emission is seen at $> 2 \sigma$ in any of the flux-density stacked maps of the sources. Several noise peaks are seen at several arcseconds removed from the centre of the stack. To confirm these as random noise fluctuations, visual inspection of the distribution of fluxes across the source (i.e., flux density histograms) suggest that it is Gaussian without emission seen at high or low signal-to-noise. This assuages any concerns regarding the combination of multiple maps with different beam sizes, which are not convolved to the same beam size to conserve the observational depth. This is in line with the algebraic expectation, as the stacking procedure in this paper is a linear combination of maps with Gaussian noise profiles, which should conserve the Gaussian noise properties regardless of the final, much more complicated beam shape.

\subsection{Stacking to a single dust mass}
\label{sec:stackingtoasingledustmass}
In order to combine observations at different rest-frame frequencies into a single representative image, we can use a single-temperature optically-thin modified blackbody template to scale the observations to a single dust mass. To this end, we convert the individual images from Jy/beam to $M_{\rm dust}$/beam. Subsequently, the images are stacked using the same stacking method as detailed in Section~\ref{sec:sensitivityweightedstacks}, providing the highest-fidelity measure of the dust mass of an archetypal galaxy in the $z > 8$ Universe.

Figure~\ref{fig:SED} shows the relative contribution of each of the sources to a combined dust mass estimate through stacking. 
The observed fluxes of each source (shown as $3 \sigma$ limits for all sources but Y1; see Table~\ref{tab:observationProperties}) is shown against an ad-hoc redshift $z = 9$ modified blackbody with a $10^6$~M$_{\odot}$ dust reservoir at 30, 50 and 70~K. Below, we show the steps to calculate the dust mass estimate from each observation, with the results displayed in Table~\ref{tab:Dustlimits}, accounting for gravitational lensing where appropriate. Furthermore, an additional measure of the dust mass is provided through the UV attenuation, which is used for comparison against direct stacking measurements.

\subsubsection{Infrared-based dust mass estimates}
In this study, we presume the dust to be optically-thin in the infrared-observed wavelengths. The deep upper limits from even individual observations imply low dust masses, in line with the optically-thin scenario. As a sanity check, we ensure that the optically-thick wavelength implied by the dust masses and (optical) sizes is well below the rest-frame wavelength using equation 3 in \cite{Algera2024REBELS25}. As such, the following equation is used to calculate the per-beam dust mass, 
\begin{equation}
S_\nu = \left(\frac{1+z}{d_\mathrm{L}^2}\right) M_\mathrm{dust} \kappa_0 \left(\frac{\nu}{\nu_0}\right)^{\beta_\mathrm{dust}} \left[B_\nu(T_{\mathrm{dust},z}) - B_\nu(T_{\mathrm{CMB},z}) \right].
\end{equation}
In this equation, $d_\mathrm{L}$ is the luminosity distance, $\kappa_0$ is the dust mass absorption coefficient at $\nu_0 = 1900$~GHz, which in our case is assumed to be 10.41~cm$^2$~g$^{-1}$ based on \cite{Weingartner2001}. The dust at redshift $z$ is heated by the CMB. Throughout this work, we assume the dust temperature to be 50~K at $z = 0$, and use equation~12 from \cite{daCunha2013} to account for the additional dust warming of the CMB. This temperature is representative of galaxies in the $z > 7$ Universe \citep{Liang2019,Bakx2021,Bouwens2022REBELS}, although it can vary by up to $\sim 30$~K from source-to-source. Additionally, we include the contrast to the CMB against which we observe the dust emission at redshift $z$ through the $B_\nu(T_{\mathrm{CMB},z})$ term. In line with lower-redshift detected objects up to $z = 8$, we assume a dust emissivity index of $\beta_{\rm dust} = 2$ \citep{Bakx2021,Witstok2023DustStudy,Algera2024REBELS25}.

\subsubsection{UV-based dust mass estimates}
The rest-frame UV emission is affected by the dust. In an effort to understand the infrared-based dust mass limits, we also evaluate the dust masses implied by the galaxies' UV emission. Although a full characterisation of the dust emission requires radiative modeling, assuming a dust geometry, preliminary estimations to the total dust mass can be made. Following the formalism in \cite{Ferrara2022REBELS}, we estimate the total dust masses using UV-based sizes and a Meurer-based dust attenuation law using their equation 11, 
\begin{equation}
M_{\rm dust, opt} = f_{\mu} \frac{\tau_{1500}}{\tau_0} \left(\frac{r_d}{\rm kpc} \right)^2 {\rm M}_{\odot}. \label{eq:dustmassoptical}
\end{equation}
In this equation, $f_{\mu}$ is a geometrical correction assumed to be 4/3 (the value for a homogeneous dusty sphere), where it is presumed that the emitting sources (i.e,. stars) lie behind a foreground dust shield. $\tau_{1500}$ is the optical depth at 1500~\AA, with $\tau_0$ being set to the Small Magellanic Cloud value of $2.17 \times{} 10^{-8}$, with an alternative optical depth of the Milky Way at $1.09 \times{} 10^{-8}$ \citep{Weingartner2001}, and $r_d$ is the radius of the dust-emitting region (here presumed to be equal to the optical size of the source, although see \citealt{Sommovigo2025}). 
The $\tau_{1500}$ is calculated from the attenuation curve found by \cite{Meurer1999}, which correlates the UV-continuum slope $\beta_{\rm UV}$ against the attenuation, finding $\tau_{1500} = 1.99 (\beta_{\rm UV} - \beta_{\rm UV, intrinsic})$. The unattenuated UV-continuum slope, $\beta_{\rm UV, intrinsic}$, is found to be $\beta_{\rm UV, intrinsic} = -2.23$ \citep{Meurer1999}. 

These attenuation curves appear to hold for REBELS galaxies \citep{Schouws2022}, with a slightly lower unattenuated UV-continuum slope of $\beta_{\rm UV, intrinsic} = -2.63$ providing a better quality of fit on \textit{Hubble} data. However, a more recent NIRSpec Integral Field Unit (IFU) suggests higher unattenuated UV-continuum slopes of $\beta_{\rm UV, intrinsic} = -2.4 $ to $ -2.0$ \citep{Fisher2025}, with their shape more in line with a more shallow, Calzetti-like dust-attenuation curve. In our study, we adopt the value derived for higher-redshift galaxies from \textit{Hubble} observations ($\beta_{\rm UV, intrinsic} = -2.63$) for an estimate of the dust mass, especially since four sources have a $\beta_{\rm UV}$ below $-2.23$. Each of these parameters, as well as the co-spatial assumption of dust and stellar emission, come with large intrinsic uncertainties. This dust mass should thus instead be seen as an order-of-magnitude estimate, where large discrepancies between this UV-attenuation dust measurement and the far-infrared emission can indicate unusual dust geometries and/or dust composition properties. 

Neither REBELS-24, GS-z9-3 nor SPT-0615-JD have a measure for their optical size. 
For GS-z9-3, the $\beta_{\rm UV}$ is not available, but the DAWN JWST Archive (DJA) and its subsequent analysis provide physical estimates of $A_V = 0.29$ and $L_{\rm IR} = 6.9 \times{}10^{10}$~$L_{\odot}$ \citep{Heintz2025}. The strong lensing of SPT-0615-JD also prohibits a single size, particularly since it is revealed to contain several stellar clusters \citep{Bradley2024,Adamo2024}. For REBELS-24 and GS-z9-3, we adopt an ad-hoc size of 100~pc, in line with other sizes used for optical dust analysis in the REBELS sample \citep{Ferrara2022REBELS}, as well as other \textit{JWST}-observed galaxies at $z > 8$. Similar to the combined limit on the sub-mm dust mass stack, we estimate the average optically-derived dust mass using the sub-mm derived dust mass limits as a weighting for the optical dust masses. This way, the sub-mm derived dust mass limits combine into an average optically-derived dust mass for a typical galaxy across the sample.

\begin{figure}
    \centering
    \includegraphics[width=\linewidth]{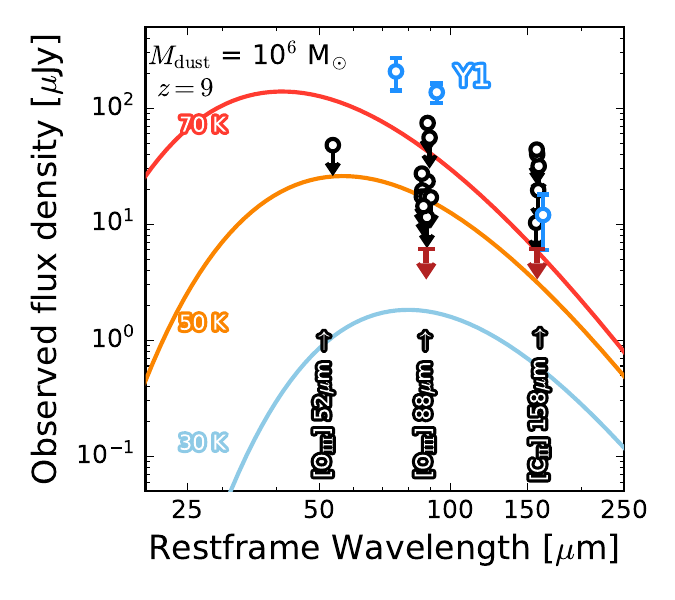}
    \caption{The observations of $z > 8$ galaxies typically target the expected wavelengths of bright sub-mm spectral lines (\oiii{} 52~$\mu$m, \niii{} 57~$\mu$m, \oiiil{}, and \ciil{}), and subsequently probe different parts of the dust spectrum. The observational limits and detections of nineteen dust continuum observations across thirteen $z > 8$ sources are shown against the expected dust emission of a $z = 9$ galaxy with a dust mass of $10^6$~M$_{\odot}$ with a $\beta_{\rm dust} = 2$ at three different dust temperatures of $T_{z = 0} =$ 30, 50 and 70~K assuming a modified black-body. The single source with dust continuum detections at $z > 8$, Y1, is shown in blue. The stacks at rest-frame 88~ and 158$\mu$m are shown as red arrows.}
    \label{fig:SED}
\end{figure}

\subsection{Stacking to a single dust-to-stellar mass ratio}
\label{sec:stackingtoasingledusttostellarmass}
Since more massive galaxies likely also have larger dust masses, we also perform a stack to test the limits in dust-to-stellar mass ratios from recent observations. Using the published stellar masses derived from the integrated rest-frame UV- and optical photometry, we calculate the dust-to-stellar mass ratio in a per-pixel fashion prior to stacking. Here, we note the important caveat that the diversity in spectral fitting approaches can result in systematically different stellar masses across the sample. 
The effect of gravitational lensing is presumed to affect the stellar and dust components equally, ignoring any effects of differential lensing \citep{serjeant2012}, and is thus effectively negated in the per-pixel measure of the dust-to-stellar mass ratio. The stacking weight of each image is based on the standard deviation in dust-to-stellar mass ratio.

\subsection{Binning by redshift, $\beta$ and stellar mass}
Although this sample of ten galaxies is modest, it allows us to test different subsets of populations against each other. In total, we chose criteria to separate the population of galaxies, namely redshift, stellar mass and UV spectral slope ($\beta_{\rm UV}$). Although more binning options are imaginable, these result in a higher chance of false positives through \textit{p-hacking}, and the scientific motivation behind these three bins appears most robust. 

Firstly, dust formation likely occurs on timescales similar to or in excess of the age of the Universe at $z \sim 8 - 14$. %Consequently, the amount of dust is expected to be higher for lower-redshift galaxies in this range, presuming negligible dust destruction effects in the observations. 
In an effort to see the effect of additional available time on the formation of dust, we test the binning on a low- and high-redshift bin, roughly taking the median redshift of the sample at $z = 9.5$ to classify these bins. 

Secondly, galaxies with larger stellar masses are expected to have both larger total gas masses, and higher gas-phase metallicities \citep[e.g.,][]{Nakajima2023}, both of which result in higher dust masses at constant dust-to-metal ratio. As well, models \citep[e.g.,][]{Feldman2015,Esmerian2022,Esmerian2024} predict that for the most plausible dust physics parameters, the dust-to-metal ratio depends positively on metallicity \citep{Hirashita2012,Dayal2022}, which is also observed in galaxies in the local Universe \citep{Remyruyer2014,DeVis2019,Galliano2021} albeit with substantial scatter. In an effort to see the effect of stellar mass on the formation of dust, we test the binning on a low- and high-stellar mass bin, taking roughly the median mass of the sample at $> 10^9 \,\rm M_{\odot}$ to classify these bins. 

Finally, the dust attenuation of a system is reflected directly in the UV-continuum slope $\beta_{\rm UV}$. While a $\beta_{\rm UV}$ below $-2$ indicates a relatively dust-poor system, obscured -- and thus dust-rich -- systems typically have higher $\beta_{\rm UV}$ (e.g., $> -2$ \citealt{Mitsuhashi2024}), although spatial offsets between dust and UV emitting regions significantly complicate this interpretation \citep{Sommovigo2022}. Subsequently, we test the binning on a low- and a high-$\beta_{\rm UV}$ bin, where we note that higher (less negative) $\beta_{\rm UV}$ are typically expected to be dustier \citep{Fudamoto2020,Bowler2024}. 

{\color{referee2}
\subsection{Choice of stacking method for reporting results}
\label{sec:stackingconsiderations}
Throughout this study, we employ three stacking approaches: flux-weighted (Section~\ref{sec:sensitivityweightedstacks}), dust-mass-weighted (Section~\ref{sec:stackingtoasingledustmass}), and dust-to-stellar-mass-ratio-weighted (Section~\ref{sec:stackingtoasingledusttostellarmass}). Each method has distinct advantages. The dust-mass-weighted approach achieves the deepest individual limits by optimally weighting strongly-lensed sources, but risks being dominated by a single object (e.g., MACS1149-JD1 with $\mu$ = 10.5). Conversely, the dust-to-stellar-mass-ratio approach weights all sources more equally since lensing affects both components similarly, although it introduces additional uncertainty from stellar mass measurements.

For our dust mass results (Figures~\ref{fig:MstarMdust}; left-hand side, \ref{fig:Lz} and \ref{fig:Mstar_vs_MstarMdust_stack}; left-hand side), we adopt the conservative flux-weighted approach, converting flux limits to dust masses at the average observing frequency and redshift, which are listed in the final column in Table~\ref{tab:stacking}. The flux-based dust mass estimate is about a factor five times more conservative than the dust-stacked mass estimate, but better samples our full galaxy population, avoiding over-weighting of individual lensed systems while still benefiting from the improved sensitivity of stacking (see e.g., Fig~\ref{fig:SED}). Meanwhile, the dust-to-stellar mass estimates do not strongly favour lensed sources, and we believe it is best to report these results directly from the dust-to-stellar mass stacks in Figures~\ref{fig:MstarMdust} (right-hand side), \ref{fig:z_vs_MdustMstar} and \ref{fig:Mstar_vs_MstarMdust_stack} (right-hand side).
}

\section{Implications}

\subsection{Dust mass limits at $z > 8$}
Table~\ref{tab:stacking} shows the results of the dust mass stacking experiments, together with the binning in different redshift, stellar mass and $\beta_{\rm UV}$. The images of the stacks in sub-mm flux (Figure~\ref{fig:SnuMassStacks}), dust mass (Figure~\ref{fig:DustMassStacks}) and dust-to-stellar mass ratio (Figure~\ref{fig:DustToStarMassStacks}) are shown in their respective Appendices. 
None of the stacks have led to a $> 3 \sigma$ detection, although the $\beta_{\rm UV} > -2$ (i.e., dustier UV-continuum slope) indicate a $2 \sigma$ feature in the flux and dust-to-stellar mass ratio stack. Since these are drawn from a total of 21 tests, this cannot be classed as a significant result. {\color{referee2}Throughout this section, we report dust masses derived from flux-weighted stacks (Section~\ref{sec:sensitivityweightedstacks}) to ensure our limits represent the broader galaxy population rather than being dominated by individual strongly-lensed sources.}

\begin{figure*}
    \centering
    \includegraphics[width=0.475\linewidth]{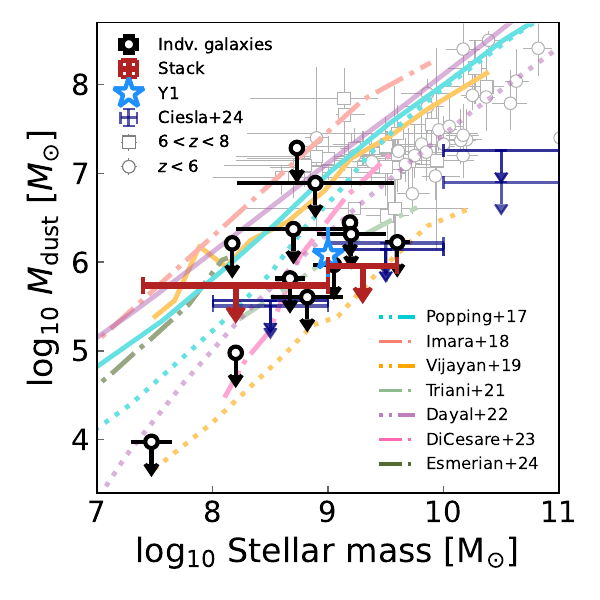}
    \includegraphics[width=0.515\linewidth]{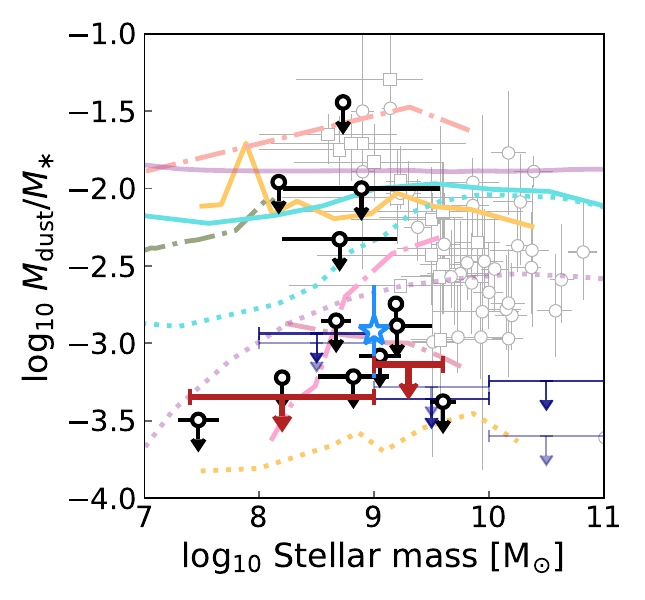}
    \caption{The dust mass and dust-to-stellar mass ratio of the individual galaxies (black circles) and of the stack at two different stellar mass bins ($\log_{10} M_{\ast}/{\rm M_{\odot}} \lessgtr 9$; red upper limits) as a function of stellar mass in the left and right panel, respectively. The upper limits are drawn at $3 \sigma$ and the dust and stellar masses are corrected for lensing. These results are compared against individually-detected galaxies at $4.4 < z < 6$ indicated with grey circles \citep{Sommovigo2022} and $6 < z < 8$ galaxies indicated with grey squares \citep{Bakx2021,Sommovigo2022REBELS,Witstok2022,Fudamoto2022DustTemperatures,Algera2024,Algera2024REBELS25,Valentino2024}. The sole source with a dust detection at $z = 8.3$ is shown as a blue star \citep{Tamura2019,Tamura2023,Bakx2020CII}. Dust estimates from stacking experiments are shown as thin blue upper limits \citep{Ciesla2024}, and scaling relations from semi-analytical and hydrodynamical models that account for dust production are shown as trend lines between $z = 7$ to $9.5$ \citep{Popping2017Dustproduction,Imara2018,Vijayan2019,DiCesare2023,Esmerian2024,Triani2021,Dayal2022}. For two dust production models \citep{Popping2017Dustproduction,Vijayan2019}, we show the maximum and average dust production scenarios in solid and dash-dotted lines respectively.}
    \label{fig:MstarMdust}
\end{figure*}

These dust non-detections beyond $z > 8$ are in stark contrast to the detections in the $5 < z < 8$ Universe. Figure~\ref{fig:MstarMdust} compares the stacked and individual dust masses and dust-to-stellar mass ratios assuming a 50~K modified black-body with a $\beta_{\rm dust} = 2$ against the stellar masses. The stacks are divided in a lower- and higher stellar mass bin at $\log_{10} M_{\ast}/{\rm M_{\odot}} \lessgtr 9$. These are compared against lower redshift galaxies at $4.4 < z < 6$ from the ALPINE sample \citep{Sommovigo2022} and from REBELS and individual observations at $6 < z < 8$  \citep{Bakx2021,Sommovigo2022REBELS,Witstok2022,Fudamoto2022DustTemperatures,Algera2024,Algera2024REBELS25,Valentino2024}. Note that these measurements mostly come from a single dust continuum data point.
The single source with dust detections above $z > 8$, Y1, is also shown for comparison. 
The stacking experiment from \cite{Ciesla2024} is converted to a dust mass and dust-to-stellar mass ratio estimate assuming the same 50~K and $\beta_{\rm dust} = 2$. 
The data points are compared to scaling relations from semi-analytical models and hydrodynamical simulations that account for dust production \citep{Popping2017Dustproduction,Imara2018,Vijayan2019,DiCesare2023,Esmerian2024}. These scaling relations are taken at the redshift range similar to our stacked sample, namely at $z = 7$, 9.5, 7, 9.5 and 8.5, respectively. For the \cite{Popping2017Dustproduction} and \cite{Vijayan2019} dust production simulations, we show the maximum and average dust production scenarios that assume differences in supernova dust production rates and dust condensation efficiencies. 

The dust masses and dust-to-stellar mass ratios of individual galaxies and the stacks lie below the more prolific dust productions, providing a constraint on the dust production physics included in models. Importantly, many of the galaxy samples do not include the upper limits, and the build-up represented by the detections shown in grey could be biased high in Fig.~\ref{fig:MstarMdust}. Regardless, at $z > 8$ only one source (Y1; \citealt{Tamura2019}) provides a dust detection. We note that its relatively low dust mass (and dust-to-stellar mass ratio) is due to a high dust temperature (91~K; \citealt{Bakx2020CII}, \citealt{Bakx2025}), which only recently has been confirmed by short-wavelength observations. 

% What is the effect of dust temperature variation
The dust masses inferred from the individual observations and stacks in this study presume constant dust properties, including a fixed dust temperature ($T_{z = 0}=50$~K) and $\beta_{\rm dust}$ (= 2). If the assumed dust temperature is changed by $\pm 10$~K, the predicted dust masses across the entire sample changes roughly by a factor of two, with the colder dust temperature predicting higher dust masses and vice versa. If the assumed dust emissivity index $\beta_{\rm dust}$ is changed by $\pm 0.5$, the predicted dust masses across the entire sample changes by roughly 25 per cent, with lower $\beta_{\rm dust}$ predicting higher dust masses and vice versa. These results are in line with the analytical analysis presented in \cite{Algera2024REBELS25}. 
Similarly, in the fitting the dust is presumed to be optically thin. This effect is thought to be minor, as dust masses are necessarily low -- since we are turning to stacking -- and the effect of the optical depth will likely not play a significant role in our dust mass estimates. The dust temperature distribution, as well as diverse between sources, is likely also diverse within these galaxies. Recent analytical modeling \citep{Sommovigo2025DifferentTdistributions} predicts under-estimation of the dust mass up to 0.25~dex at $T_{z = 0} = 50$~K for realistic internal dust temperature variations  $\sigma_{T} / \bar{T} < 0.3$, in line with resolved observations of dust temperature variations in $z = 7$ galaxies \citep{Akins2022}.

Figure~\ref{fig:SED} shows the relative contribution of the rest-frame 52, 88 and 158~$\mu$m to the deep dust mass limits.
Observations at rest-frame 88 and 158~$\mu$m are equally sensitive to a cold dust temperature (30~K), but the primary measure of the dust mass of a 50~K body occurs at rest-frame 88~$\mu$m and shorter. Counter-intuitively, we find a deeper limit for the 158~$\mu$m than 88~$\mu$m, which is due to more lensed targets in the 158~$\mu$m sample compared to the 88~$\mu$m sample. An evolution of the dust temperature with redshift (e.g., \citealt{Liang2019,Sommovigo2021}) would mean that shorter wavelength observations ($\leq 88$~$\mu$m in rest-frame wavelength) are favoured more in dust mass estimates. Without direct observational arguments towards a dust temperature evolution beyond $z > 8$, the most conservative estimate is a constant dust temperature of $T_{z = 0} = 50$~K, although it is important to note that any evolution pushing for much higher dust temperatures would provide even deeper constraints on the dust production physics at $z > 8$.

Meanwhile, $\beta_{\rm dust}$ can vary between sources \citep[e.g.,][]{Witstok2023DustStudy}. A lower $\beta_{\rm dust}$ would imply a less steep Rayleigh-Jeans long-wavelength tail, which would increase the relative sensitivity of the longer-wavelength data points on the total dust mass. Conversely, a higher $\beta_{\rm dust}$ -- as indicated in several high-redshift galaxies \citep{Algera2024REBELS25} -- would have a similar effect as a warmer dust temperature. This $\beta_{\rm dust}$ and dust temperature degeneracy can complicate studies with multiple data points \citep[e.g.,][]{Bakx2020CII}, although the lack of sample-wide variation in the $z > 7$ Universe suggests this is only a modest effect ($\sim 25$~per cent) within our study.

% What is the effect of source-to-source variation
The upper limits on the dust mass are determined mostly by the sources with deep intrinsic observations. In particular, ID4590 and MACS1149-JD1 are both lensed by $\mu \approx 10$, resulting in strong estimates of the total dust mass. Since galaxies with lower stellar masses are predicted to have significantly smaller dust masses due to inefficiencies in the interstellar grain growth, the higher stellar mass bins provide a substantially improved constraint on the measure of dust production when compared to dust models \citep{DiCesare2023}.

{\color{referee} Our work is a natural continuation of stacking exercises at cosmic noon pushed out to higher redshifts. Although a direct comparison of galaxies at cosmic noon and at $z > 8$ is complicated by the different metallicities, dust composition and production/destruction pathways \citep{Boquien2022,Witstok2023,Markov2023,Sommovigo2025}, the much larger spectro-photometric catalogues also provide deep measures and limits on the dust-to-stellar mass ratio at $z = 1.5 - 5$ \citep{Bouwens2016,Bouwens2020,Dunlop2017,McLure2018} as well as at $z = 6 - 8$, where sample-wide estimates exist from surveys such as from REBELS \citep{Algera2023,Bowler2024}. Furthermore, samples with robust metallicity estimates now offer a chance to investigate the effect of metallicity on the dust production \citep{Shivaei2022}. In Appendix Figure~\ref{fig:Mstar_vs_MstarMdust_stack}, we compare these reference samples against the deep upper limits from this study, as well as against the $z = 6 - 8$ scaling relations detailed in Figure~\ref{fig:MstarMdust}. For all cosmic noon studies where dust masses were not explicitly given, we assume a 35~K dust temperature and a dust-emissivity index $\beta_{\rm dust}$ of $2$. Even though some studies at cosmic noon only find upper limits after combining hundreds or even thousands of galaxies, these values are in line with the dust scaling relations at $z = 6 - 8$, similar to the $z \sim 2$ solar- and lower-metallicity galaxies studied in \cite{Shivaei2022}. The deep limits on the dust-to-stellar mass ratio from our $z > 8$ study still stand out relative to the individual and composite data at $z < 8$, as well as the dust scaling relations, due to the combined effect of K-correction and the likely higher dust temperatures at $z > 5$.  }

% Mdust_optical
The UV-continuum slope provides indications of dust-obscuration in the $z > 8$ Universe, with an average of $\beta_{\rm UV} = -2.1$ across the sources that are stacked in line with the average of the REBELS sample ($\beta_{\rm UV} = -2.0$; \citealt{Bouwens2022REBELS}). Although full radiative transfer modeling, including accurate estimates of the attenuation curves (c.f., \citealt{Markov2023}), the geometry \citep{casey2014,dunne2018unusual,Vijayan2025}, and the unattenuated stellar emission ($\beta_{\rm UV,0}$; see e.g., \citealt{Fisher2025}), is necessary to provide an estimate of the total dust mass from this observed quantity, we provide a rough estimate of the total dust mass through Equation~\ref{eq:dustmassoptical}. 

\begin{figure}
    \centering
    \includegraphics[width=\linewidth]{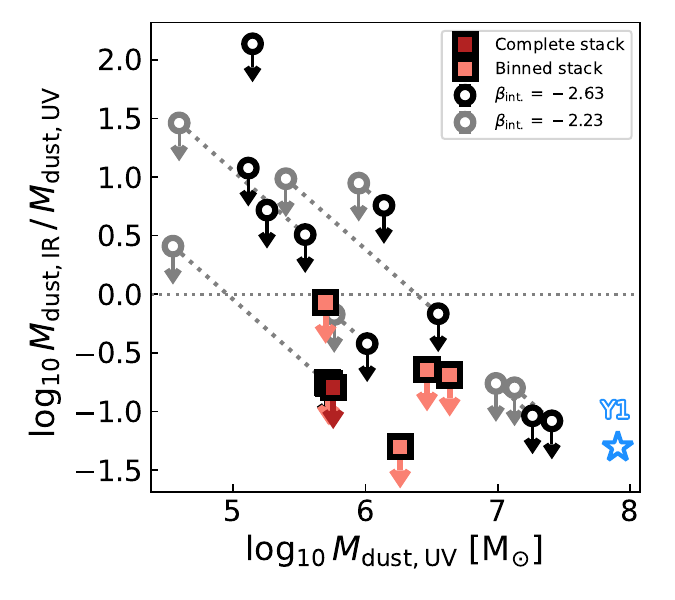}
    \caption{The UV-derived dust mass against the infrared to UV-derived dust masses for our sample and stack. The difference between the UV- and infrared-derived dust masses indicate the limitations of the UV-based perspective on dust production and obscured star formation. The dark circles indicate the optically-derived upper limits based on $\beta_{\rm UV, intrinsic} = -2.63$, while the connected grey points indicate the $\beta_{\rm UV, intrinsic} = -2.23$ scenario as suggested for lower-redshift galaxies \citep{Meurer1999}. Note that these are unavailable for sources with $\beta_{\rm UV} < -2.23$. Individual sources, as well as all the stacks (dark red for the complete stack, with orange colours indicating the binned stacks) lie on the side where the UV-derived dust masses are $\sim 1$~dex in excess of infrared-derived optically-thin dust masses with $T_{\rm dust} = 50$~K and $\beta_{\rm dust} = 2$. The UV-estimated dust mass for Y1 (in light blue) is also larger than the measured value \citep{Tamura2019}. }
    \label{fig:mopt}
\end{figure}

Figure~\ref{fig:mopt} compares the estimates of the total dust mass through UV and infrared measures, for both individual sources and the (binned) stacks. The UV-based dust mass estimates in the stacks are calculated using the same weighting as the infrared-based dust mass estimates, to ensure a fair comparison. Several individually-observed sources and, more importantly all of the stacks find higher optical dust masses than the sub-mm dust mass limits. While the assumption on the intrinsic $\beta_{\rm UV}$ or the sub-mm emitting size (as compared to the rest-frame UV sizes used in this study) can significantly reduce the total presumed dust mass \citep{Ferrara2024EventfulLife}, it appears that the UV sizes and continuum slope do not provide an accurate measure of the dust masses in the early Universe with the assumptions used in this work, in line with previous sources with dust continuum detected sources \citep{Ferrara2022REBELS} and Y1 \citep{Tamura2019}. This could point to spatial offsets between the dust-emitting regions and the UV-emitting regions \citep{Sommovigo2022z5galaxies}. 
Although several sources have confirmed high (i.e., dusty) $\beta_{\rm UV}$ from NIRSpec observations, nebular lines can affect the slopes in the broad continuum photometry \citep{Saxena2024,Katz2024}.
Alternatively, theories regarding dust destruction or dusty outflows \citep{Ferrara2025AFM} could affect the geometric arguments. Similarly, our assumption on the dust properties, in particular the dust temperature of $T_{\rm dust} = 50$~K, affects the total dust mass present in galaxies in the early Universe.

% UV preselection of these galaxies?
As seen in Figure~\ref{fig:MuvVsZ}, many of the galaxies in this survey are the UV-brightest objects. This UV-preselection can result in missing the most dust-obscured systems at a given redshift, as exemplified by so-called optically-dark galaxies seen out to $z > 7$ \citep{Fudamoto2021OpticallyDark}. Attempts to circumvent these pre-selections are possible through emission-line galaxy studies \citep{venemans2020,Bakx2024,vanLeeuwen2024MNRAS.534.2062V}, which find much higher dust masses in the $z > 6$ Universe. In particular given the small optical sizes of these galaxies where little dust is needed to redden their UV-colours to below detection limits (see eq.~\ref{eq:dustmassoptical}), the dustiest galaxies at a given epoch could be missed by JWST-selected samples. Fully excluding large dust reservoirs in the $z > 8$ Universe thus also requires future updates to the mapping speed of ALMA \citep[e.g.,][]{carpenter2022alma2030} and the advent of future sub-mm single-dish institutes \citep{vanKampen2024} to test dust production across cosmic time. 

\subsection{Dust build-up in the early Universe}
% The dust build-up timescales are comparable to the age of the Universe at $z = 8$. However, the deep upper limits found in this study, both in individual observations and through stacks, highlight that we might be approaching the limits in our quest to find the advent of cosmic dust. 

\begin{figure}
    \centering
    \includegraphics[width=\linewidth]{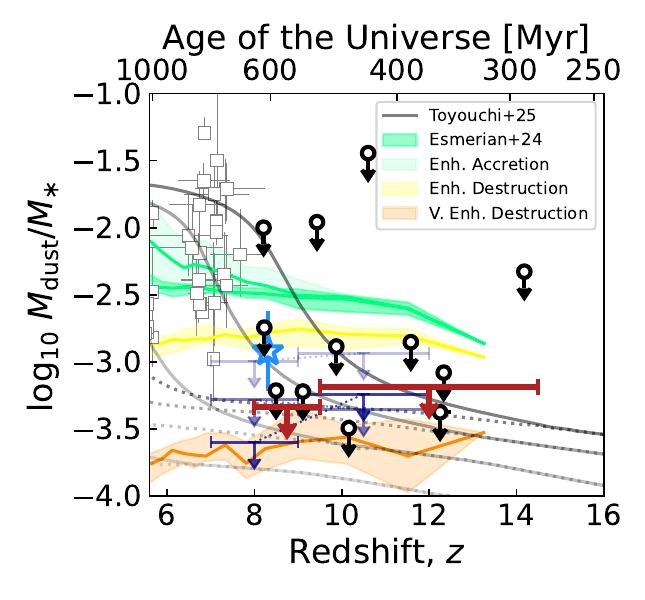}
    \caption{The dust-to-stellar mass ratio as a function of redshift and cosmic time. All $z > 8$ individual sources (\textit{black points}; except Y1 in \textit{blue}) remain undetected, as are the stacked images (\textit{red squares}). Blank-field stacking experiments (\textit{dark-blue downward triangles} with the darkest-to-lightest hue corresponding to $M_{\ast} > 10^{10} $~M$_{\odot}$, $> 10^{9} $~M$_{\odot}$, and $> 10^{8} $~M$_{\odot}$, respectively; \citealt{Ciesla2024}). Individual dust detections (grey squares; \citealt{Bakx2021,Sommovigo2022,Sommovigo2022REBELS,Witstok2022,Fudamoto2022DustTemperatures,Valentino2024}) suggest rapid and efficient dust growth in the $z < 8$ Universe. The solid and dashed lines indicate efficient and inefficient grain growth timescales ($\tau_{\rm ISM\,growth} = 5$ and 50~Myr, resp.) for different dark-matter halos with $z = 5$ masses of $\log_{10} M_{\rm halo}/$M$_{\odot}$ of 10, 11, 12 and 13 with increasing hue \citep{Toyouchi2025}. The filled regions indicate the dust-to-stellar mass ratio evolution from a hydrodynamical model \citep{Esmerian2024} across different dust prescriptions, including (very) enhanced dust accretion and destruction. }
    \label{fig:z_vs_MdustMstar}
\end{figure}

Figure~\ref{fig:z_vs_MdustMstar} shows the dust-to-stellar mass ratio for our galaxies and stacks as a function of redshift. The expected dust-to-stellar mass ratio is compared to other stacks \citep{Ciesla2024}, detected galaxies \citep{Bakx2021,Sommovigo2022,Sommovigo2022REBELS,Fudamoto2022DustTemperatures,Witstok2022}, an analytical one-dimensional dust evolution study \citep{Toyouchi2025} and the redshift evolution of the dust-to-stellar mass ratio seen in a hydrodynamical model \citep{Esmerian2024}. As in the dust mass and dust-to-stellar mass ratio study, we convert the flux density upper limits of \cite{Ciesla2024} to a dust mass using a 50~K optically-thin dusty body with $\beta_{\rm dust} = 2$. The dust evolution study provides estimates of metal build-up in a self-consistent, radially-resolved galaxy evolution model for galaxies hosted in different dark-matter halos with $z = 5$ masses of $\log_{10} M_{\rm halo}/$M$_{\odot}$ of 10, 11, 12 and 13. Subsequent dust production is then predicted through a metal accretion timescale, i.e., interstellar grain growth timescale ($\tau_{\rm ISM\,growth}$), which is varied between accretion timescales of 5 and 50~Myr. Similarly, the fluid-dynamical model of \citet{Esmerian2024} demonstrates the effect of different dust accretion and destruction prescriptions. 

The dust-to-stellar mass ratio measures of individual sources and stacks lie below typically-observed values in the more nearby Universe at $z < 8$ of $\log_{10} M_{\rm dust} / M_{\ast} > -3$, with the only source at $z > 8$ in line with $\log_{10} M_{\rm dust} / M_{\ast} = -3$. Likely, the low efficiency of growing interstellar dust or the late AGB phase in the interstellar medium might result in low dust-to-stellar mass ratios, and require a transition in dust formation efficiency to predict the dust masses seen in the $z < 8$ Universe. {\color{referee} A first-order comparison between galaxies at $z > 8$ and those at $z \approx 7$ appears justified based on similar number densities. The bright end of the UV luminosity function shows weak redshift evolution \citep{Bouwens2021}, and recent results suggest that galaxies at $z > 8$ with comparable UV magnitudes ($M_{\rm UV} \approx 20$) exhibit similar number densities per magnitude and volume \citep{Donnan2024}.} The dust-to-stellar mass ratio model by \cite{Toyouchi2025} finds higher dust-to-stellar mass ratios for more massive dark matter halos, suggesting the average $z > 8$ galaxy with deep ALMA observations is not hosted in a halo with a higher $z = 5$ mass than $\log_{10} M_{\rm halo}/$M$_{\odot} > 12$ in the efficient dust production model that assumes a shorter ISM dust growth timescale ($\tau_{\rm ISM\,growth} = 5$~Myr). 

Alternatively, non-detections may not solely reflect inefficient dust production, but {\color{referee} could instead indicate} rapid dust removal through galactic outflows \citep{Carniani2024EventfulLife,Ferrara2025AFM}. Strong feedback mechanisms, such as fast winds ($>200$~km/s), can clear dust from these distant, sub-kpc galaxies within a few Myr. This dust clearing effect may be particularly pronounced at high redshift, where galaxies are more compact and exhibit higher specific star formation rates, potentially making them more susceptible to efficient dust ejection or redistribution through outflows. {\color{referee} Another reason for the deep dust-to-stellar mass limits could be related to dust grain growth pathways available in the early Universe. Several recent studies \citep{Shivaei2025,Narayanan2025} indicate shallower attenuation curves, potentially indicative of an abundance of large dust grains produced by supernovae \citep{McKinney2025Supernovae}. Because larger grains are less efficient at absorbing stellar light, they also produce weaker emission in the ALMA continuum bands per unit of dust mass, although detailed studies are necessary to explore the extent of this effect on high-redshift dust studies.}

Models and observations of bright $z > 8$ galaxies predict that they reside in massive cosmic overdensities \citep{Chiang2013,Chiang2015,Overzier2016,Helton2024,Witstok2025CosmicOverdensities}, where the enhanced galaxy merger rates play an important role in the necessary enrichment of the inter-stellar medium (ISM) to the rapid build-up of dust. In a similar fashion, only the very enhanced dust destruction perscription in the fluid-dynamical model of \cite{Esmerian2024} predicts similar dust-to-stellar mass ratios to those observed in most individual and stacked ratios in this study. Dust production thus appears to be inefficient in the $z > 8$ Universe, and needs to rapidly become more efficient in the $z < 8$ Universe to match the dust masses observed at $z < 8$ \citep[c.f.][]{Algera2025}.

\subsection{The present and future of dust studies in the $z > 8$ Universe}
\begin{figure*}
    \centering
    \includegraphics[width=\linewidth]{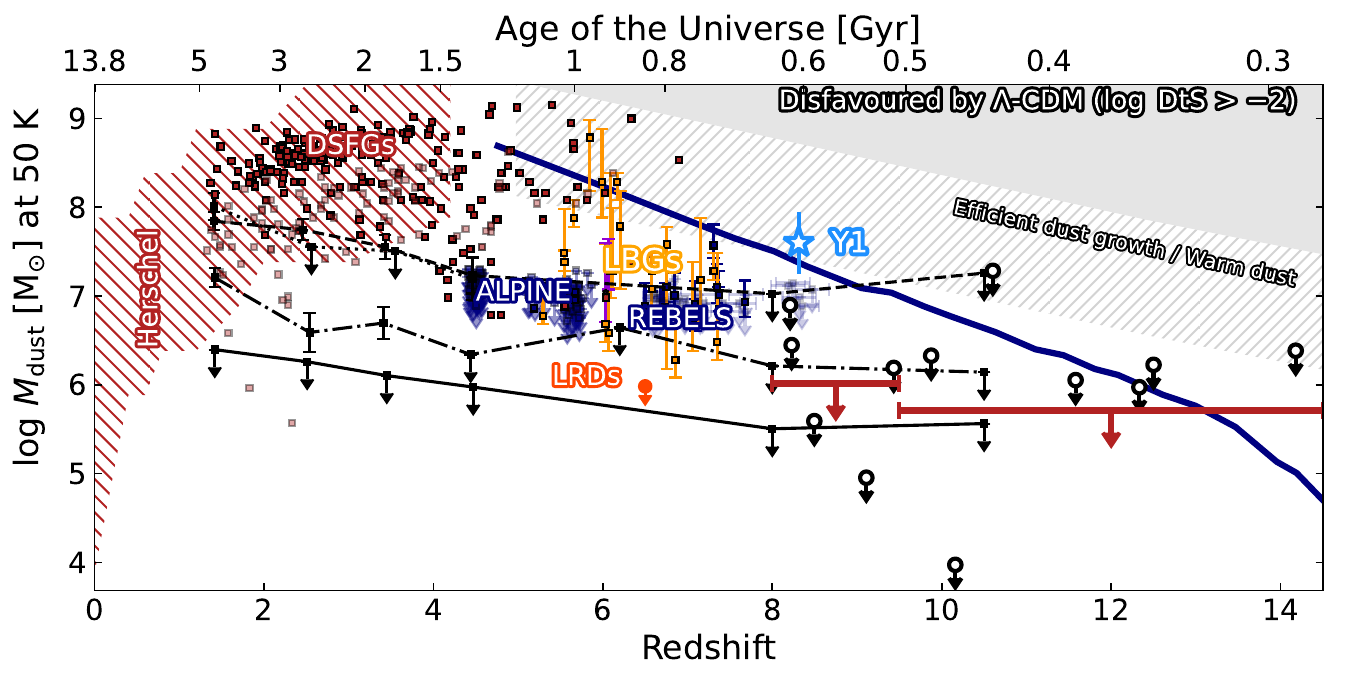}
    \caption{The dust masses from our deep sub-mm observations (\textit{black circles}), stacks (\textit{red limits}), and MACS0416\_Y1 (\citealt{Bakx2025}; \textit{light blue star}) are compared against the literature assuming $T_{\rm dust} = 50$~K. Lower-redshift Lyman-break galaxies at $z = 5 - 8$ \citep[\textit{yellow squares}][]{Laporte2019,Faisst2020,Harikane2020,Sugahara2022ApJ...935..119S,Mitsuhashi2024}, and UV-detected galaxies targeted in the ALMA Large Programs ALPINE and REBELS \citep[\textit{blue squares}][]{LeFevre2020,Bethermin2020,Bouwens2022REBELS,Inami2022}, including the upper limits identified from ALMA non-detections. Quasar-companion galaxies (QCGs) are identified from the large-area surveys \citep[\textit{purple squares}][]{venemans2020,Bakx2024}. The deep stack of sixty ``Little Red Dots'' is shown in orange \citep{Casey2025LRD}.
    % low-z
    The lower-redshift infrared sources are identified from large-area surveys at sub-mm (\textit{Herschel Space Observatory} and JCMT/SCUBA-2) and mm (\textit{South Pole Telescope}) wavelengths \citep{berta2011,Gruppioni2013,Riechers2013,Zavala2018,reuter20,ismail2023z}.
    % stacking 
    Stacking studies \citep{Ciesla2024,jolly2025} measure of the dust masses of galaxies with across a spectrum of stellar masses ($M_{\ast} > 10^{11} $\,M$_{\odot}$, $M_{\ast} > 10^{10} $\,M$_{\odot}$, $M_{\ast} > 10^{9} $\,M$_{\odot}$, $M_{\ast} > 10^{8} $\,M$_{\odot}$ in \textit{dotted, dashed, dash-dotted} and \textit{solid lines}). Most tests result in an upper limit in the dust mass, and associated infrared luminosity assuming T$_{\rm dust}$ = 50~K. 
    % scaling relations
    A comparison to the galaxy evolution model of \citet{Behroozi2018} provides a conservative estimate of the maximum infrared luminosity for a galaxy detected in a field observed by deep optical observations ($A_{\rm survey} = 0.2$ sq. deg.), assuming a complete conversion of a high baryonic-to-dark matter ($M_{\rm baryon} / M_{\rm halo} = 0.1$) to stars. The predicted limits are estimated using conservative dust-to-stellar mass ratios of 0.01 (disfavoured by $\Lambda$-CDM; \textit{solid grey region}) and of 0.001 (efficient dust growth / warm dust; \textit{hatched grey region}), as well as a realistic stellar-to-dark matter halo mass ratio \citep{behroozi2020} with an assumed stellar-to-halo mass of 0.01 (\textit{solid blue line}).}
    \label{fig:Lz}
\end{figure*}

Since the launch of \textit{JWST} and the subsequent rapid follow-up with institutes such as ALMA, the observational frontier of dust studies is rapidly evolving in the $z > 8$ Universe. In this subsection, we evaluate the present and future scope of dust mass measurements across the redshift 0 to 15 Universe based on the results from targeted observations, surveys, stacks, and simulations. 

Figure~\ref{fig:Lz} compares the $z > 8$ dust limits against estimated dust masses of galaxies, stacks and models in the $z = 0$ to $z = 15$ Universe assuming 50~K and $\beta_{\rm dust} = 2$. Lyman-break galaxies at $z = 5 - 8$ have been detected \citep{Laporte2019,Faisst2020,Harikane2020,Sugahara2022ApJ...935..119S,Mitsuhashi2024}, starting with the stunning discovery of dust at $z = 7.13$ in A1689-zD1 in 2015 \citep{watson2015dusty,knudsen2017merger,Bakx2021,Akins2022}. Since 2017, the ALPINE and REBELS ALMA large programs have aimed to characterize the dust emission of UV-detected galaxies, with success rates between 30 and 50 per cent \citep{LeFevre2020,Bethermin2020,Bouwens2022REBELS,Inami2022}. These UV-identified objects studied by these large programs appear to have massive dust reservoirs \citep{Algera2024}, and perhaps as a consequence of this, much lower dust temperatures \citep{Sommovigo2022REBELS}. 

At redshifts below 7, galaxies have been identified through their dust emission directly. Infrared surveys, starting with the mapping of the Hubble Deep Field in 1997 \citep{Smail1997,Hughes1998}, have identified over a million objects across fields with tens of square degrees to the entire sky in the sub-mm (e.g., \textit{Herschel Space Observatory} and JCMT/SCUBA-2) and mm (e.g., \textit{Planck}, {South Pole Telescope}, {Atacama Cosmology Telescope}). Consequently, these surveys readily detect sources with (apparent) luminosities beyond $> 10^{12.5}$\,L$_{\odot}$, particularly in the $z = 1 - 5$ region, with a steep drop-off at redshifts approaching zero \citep{berta2010,berta2011,Gruppioni2013,Riechers2013,Zavala2018,Magnelli2020,reuter20}. The dust temperatures of these objects are low ($\sim 30$~K), but the vast dust reservoirs ($\sim 10^{9 - 10}$~M$_{\odot}$) of these systems \citep{ismail2023z,bendo2023} dominate the extragalactic infrared emission.

Quasar-companion galaxies (QCGs), detected through line and/or continuum emission, provide an important third probe between the UV- and sub-mm selection of galaxies in the early Universe \citep{venemans2020,Bakx2024,vanLeeuwen2024MNRAS.534.2062V}. Identified using their \cii{} emission from large-area surveys, they provide a roughly star-formation rate selected sample \citep{deLooze2014}. They appear to have low dust temperatures, large dust masses \citep[$\sim 10^8\,$M$_{\odot}$][]{Bakx2024}, and high UV dust attenuations \citep{vanLeeuwen2024MNRAS.534.2062V}, with infrared luminosities in line with the LBG population.
The near-infrared capabilities of \textit{JWST} have enabled the identification of numerous compact (i.e., mostly unresolved) objects with red rest-frame optical colours. A deep stacking study on these "Little Red Dots" indicate small dust masses \citep{Casey2025LRD}. 

Stacking studies at cosmic noon (using the ALMA Lensing Cluster Survey -- ALCS; \citealt{jolly2025}) and dawn \citep{Ciesla2024} provide a measure of the dust masses of galaxies across a spectrum of stellar masses from $M_{\ast} > 10^{11}$ to $M_{\ast} > 10^{8} $\,M$_{\odot}$. {\color{referee} As can be seen in the limits from Appendix Figure~\ref{fig:Mstar_vs_MstarMdust_stack}, these and other cosmic noon \citep{Bouwens2016,Bouwens2020,Dunlop2017,McLure2018,Shivaei2022} and dawn \citep{Algera2023,Bowler2024} stacking experiments offer deep measures and limits on the dust mass. As these studies move towards higher redshifts, the combined effect of K-correction and the likely higher dust temperatures at $z > 5$ result in deeper dust mass limits at the higher redshift regime. }  

In an effort to show the maximum expected dust masses as a function of redshift, we use the galaxy evolution model of \citet{Behroozi2018}. This simulation provides the maximum expected galaxy mass for a field of 0.2~square degrees, comparable to the deep surveys done with \textit{HST} and \textit{JWST}. This limiting dust mass should be comparable to the survey sizes for most LBGs, ALPINE, REBELS, and the $z > 8$ LBGs, selected from fields smaller than 10 degrees, but not for DSFGs and QCGs identified from \textit{much} larger area surveys. We assume a complete conversion of a high baryonic-to-dark matter ($M_{\rm baryon} / M_{\rm halo} = 0.1$) to stars. The region excluded by $\Lambda$-CDM is defined by sources with high dust-to-stellar mass ratios of 0.01, and of 0.001 (efficient dust growth / warm dust; \textit{hatched grey region} in Fig.~\ref{fig:Lz}). The \textit{solid blue line} in Fig.~\ref{fig:Lz} indicates the highest stellar masses actually expected using realistic stellar-to-dark matter halo mass ratios \citep{behroozi2020}, with an assumed stellar-to-halo mass ratio of 0.01.

The deep dust mass limits of individual $z > 8$ galaxies span a broad range of dust masses, from close to the theoretical maximum dust masses to about one or two orders of magnitude below those. The rapid increase in individual dust investigations at high redshift thus offer a good opportunity to detect dust in the $z > 8.32$ Universe. The deep stacks also probe the regime where dust detections can be expected, and the continued improvement of these stacks in the near future will likely reveal dust and its build-up in the early Universe.

Future observations of dust can both use large-area maps and targeted individual observations. Since the launch of the \textit{JWST} in late 2021, it has revealed eight of the twelve sources in this paper, and the number of robust $z > 8$ galaxies with deep dust observations is increasing rapidly. 
Comparisons of large catalogues of $z > 8$ galaxies to sub-mm data with ALMA \citep{Ciesla2024}, NOEMA \citep{meyer2024}, James Clerk Maxwell Telescope / SCUBA-2 \citep{McKinney2025} and \textit{Herschel} data \citep{Viero2022} have already identified likely lower-redshift interlopers \citep{Zavala2023Masquerading,meyer2024} and provide an important resource for future stacking exercises. 

Although bootstrapping can mitigate the effects of low-redshift interlopers across large surveys that include photometric redshift candidates, it is important to combine such large-area studies with targeted individual observations and combined stacks of well-studied objects. The use of multiple frequencies, in particular at the shorter wavelengths such as the 52~$\mu$m and 57~$\mu$m observations reported in this study, can be helpful to test the variation of the internal properties of galaxies. Systematically high dust temperatures might be one of the reasons for the many non-detections at the relatively-long wavelengths reported in this study, and these can be mitigated by deep surveys at shorter wavelengths.

One exciting opportunity for stacking studies is the COSMOS High-$z$ ALMA-MIRI Population Survey (CHAMPS) ALMA Large Program (PID: 2023.1.00180.L; P.I. A. Faisst). It will combine deep MIRI imaging of the COSMOS-Web with the stacking abilities across thousands of galaxies. Likely, this project will increase the total stacking area relative to \cite{Ciesla2024} by one order of magnitude, probing a larger population of massive galaxies and improving the observational depths by 0.5~dex. 
With the increased capabilities of ALMA in the Wideband Sensitivity Upgrade \citep{carpenter2022alma2030} era, targeted observations will push our observational frontiers on the dust-obscured Universe, and be able to test our picture of dust formation and galaxy evolution.

To facilitate these stacking efforts on sources with robust redshifts, the (sub)mm data on $z > 8$ galaxies with robust spectroscopic redshifts has been made publically available at \url{http://github.com/tjlcbakx/high-z-dust-stack}. This analysis has been made possible using a public framework using the Python-based LineStacker \citep{Lindroos2015,Jolly2020} code. Using robust candidates, this resource can provide a novel pathway towards the study of dust within the first 600 million years of the Universe. As more (sub)mm observations of high-redshift galaxies inevitably come in, this tool can be used for a community-wide search for direct dust emission, probe deeper dust-to-stellar mass ratios, and test dust formation pathways in the early Universe.

\section{Conclusions}
This paper reports a deep stacking study that combines roughly one hundred hours of ALMA and 84 hours of NOEMA on-source time across fourteen different wavelengths observed in twenty-two separate projects for ten spectroscopically-confirmed galaxies at $z > 8$. This study revealed:
\begin{itemize}
\renewcommand\labelitemi{\small \textbf{$\blacksquare{}$}}
\item a deep upper limit on the dust mass, with a typically observed galaxy containing dust reservoir smaller than $9.1 \times{} 10^4$~M$_{\odot}$, assuming a dust temperature of $T_{z = 0} = 50$~K and $\beta_{\rm dust} = 2$. Furthermore, binning the observations by redshift, UV continuum slope ($\beta_{\rm UV}$) and stellar mass does not reveal any sub-population with dust emission.
\item the low estimated dust-to-stellar mass ratios of $\log_{10} M_{\rm dust} / M_{\ast} < -3.5$ to $-3$ support inefficient dust production in the early Universe, and {\color{referee} are in line with} a rapid transition in dust production efficiencies or timescales between $z > 8$ and $z < 8$. The short cosmological timescales available limit the contributions of traditional dust production pathways, including contributions from Asymptotic Giant Branch stars and interstellar grain growth, that likely generate the larger dust reservoirs observed in the z < 8 Universe. Non-detections  {\color{referee} might also indicate} rapid dust removal through galactic outflows \citep{Ferrara2025AFM}.
\item that the optical measures of dust masses are unreliable, both for individual sources and for the stacks, in line with previous studies of detected $z < 8$ galaxies \citep{Sommovigo2022}. The large uncertainties in the dust geometry and UV attenuation laws imply that (sub)mm observations are key to test dust production mechanisms in the early Universe \citep{Vijayan2025}.
\item that the low dust mass limits are not in conflict with cosmological simulations, which similarly predict a rapid build-up of stars and therefore dust in the $z > 8$ Universe. These results encourage future stacking experiments to search for dust emission in the $z> 8$ Universe, including through the publicly-available code producing this paper \url{http://github.com/tjlcbakx/high-z-dust-stack}.
\end{itemize}

\section*{Acknowledgements}
{\color{referee}The authors kindly thank the anonymous referee for their insightful comments and suggestions to improve this manuscript.}
Financial support from the Knut and Alice Wallenberg foundation is gratefully acknowledged through grant no. KAW 2020.0081. This stacking study is made possible by the continued work on ALMA receivers in Bands~5, 6, 7, and 8 \citep{Belitsky2018,Ediss2004,Mahieu2012,Sekimoto2008}. T.H. was supported by the Leading Initiative for Excellent Young Researchers, MEXT, Japan (HJH02007) and by JSPS KAKENHI grant Nos. 23K22529 and 25K00020. AKI was supported by JSPS KAKENHI 23H00131.
Some of the data products presented herein were retrieved from the Dawn JWST Archive (DJA). DJA is an initiative of the Cosmic Dawn Center (DAWN), which is funded by the Danish National Research Foundation under grant DNRF140. The authors thank Roberto Decarli, Miroslava Dessauges, Fabian Walter, and Pascal Oesch for sharing their data on GN-z11. JW gratefully acknowledges support from the Cosmic Dawn Center through the DAWN Fellowship

%%%%%%%%%%%%%%%%%%%%%%%%%%%%%%%%%%%%%%%%%%%%%%%%%%
\section*{Data Availability}
The data in this paper are available at the GitHub page \url{http://github.com/tjlcbakx/high-z-dust-stack}. The authors want to encourage the community to work towards deep stacks of galaxies in the distant Universe, and enable the first direct detection of dust beyond $z > 8.32$.

%%%%%%%%%%%%%%%%%%%% REFERENCES %%%%%%%%%%%%%%%%%%

% The best way to enter references is to use BibTeX:

\bibliographystyle{mnras}
\bibliography{1_example} % if your bibtex file is called example.bib

% Alternatively you could enter them by hand, like this:
% This method is tedious and prone to error if you have lots of references
%\begin{thebibliography}{99}
%\bibitem[\protect\citeauthoryear{Author}{2012}]{Author2012}
%Author A.~N., 2013, Journal of Improbable Astronomy, 1, 1
%\bibitem[\protect\citeauthoryear{Others}{2013}]{Others2013}
%Others S., 2012, Journal of Interesting Stuff, 17, 198
%\end{thebibliography}

%%%%%%%%%%%%%%%%%%%%%%%%%%%%%%%%%%%%%%%%%%%%%%%%%%

%%%%%%%%%%%%%%%%% APPENDICES %%%%%%%%%%%%%%%%%%%%%

\appendix

\section{Descriptions of sources}
\label{sec:appStackedSources}

\subsection{Stacked sources}

\begin{figure*}
    \centering
\includegraphics[width=0.245\linewidth]{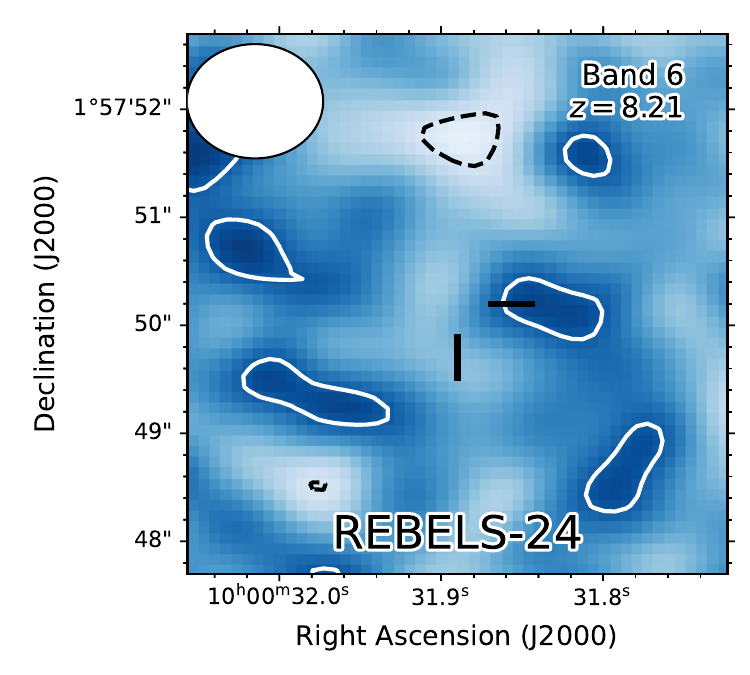}
\includegraphics[width=0.245\linewidth]{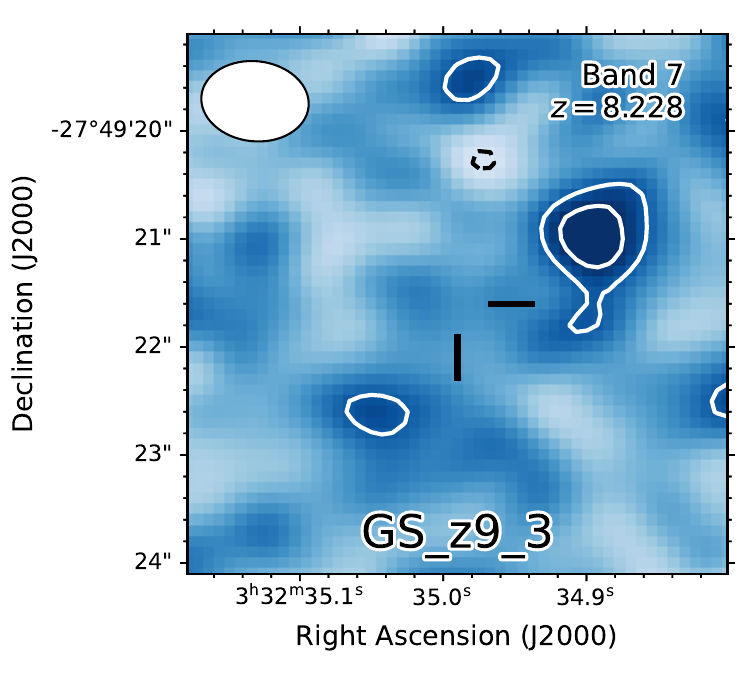}
\includegraphics[width=0.245\linewidth]{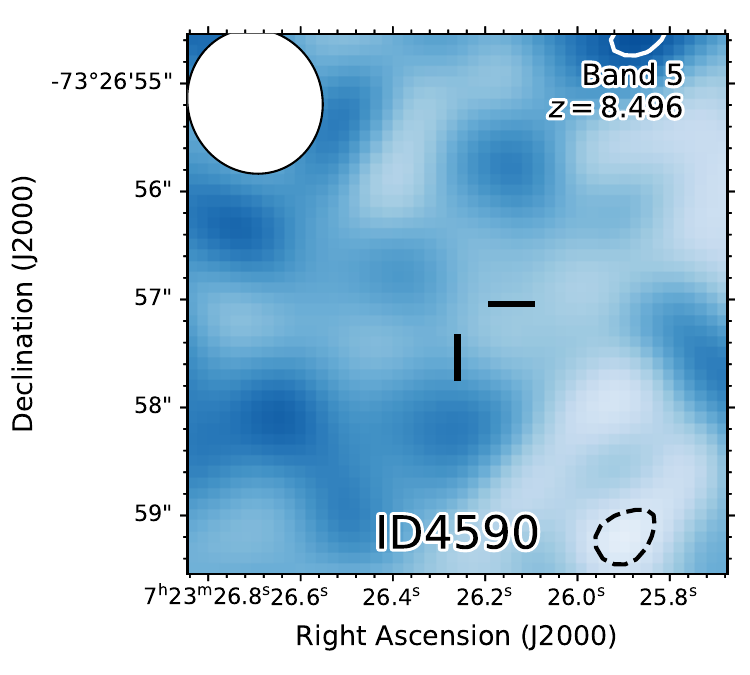}
\includegraphics[width=0.245\linewidth]{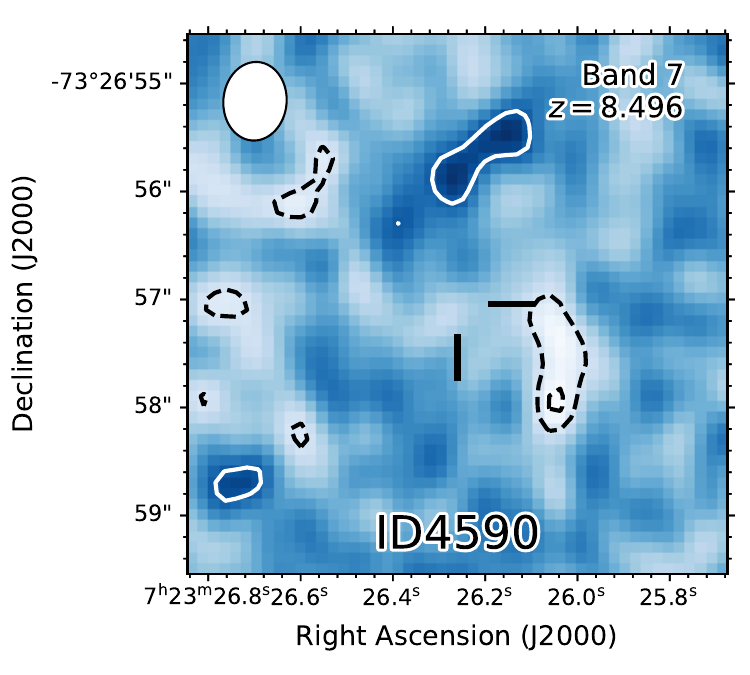}
\includegraphics[width=0.245\linewidth]{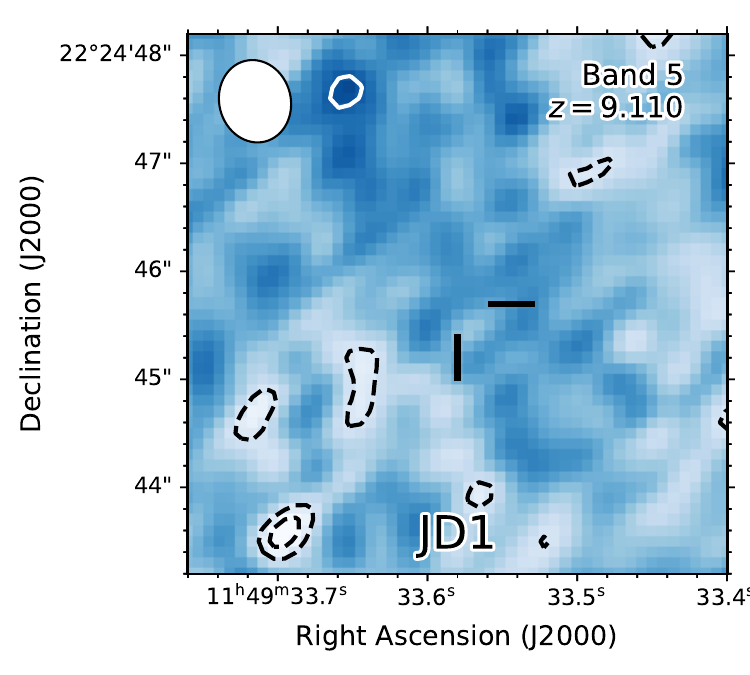}
\includegraphics[width=0.245\linewidth]{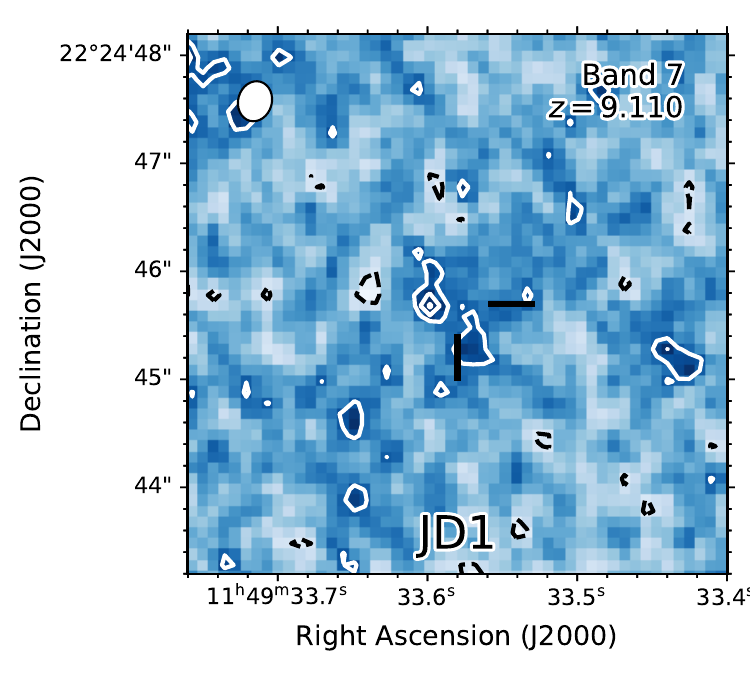}
\includegraphics[width=0.245\linewidth]{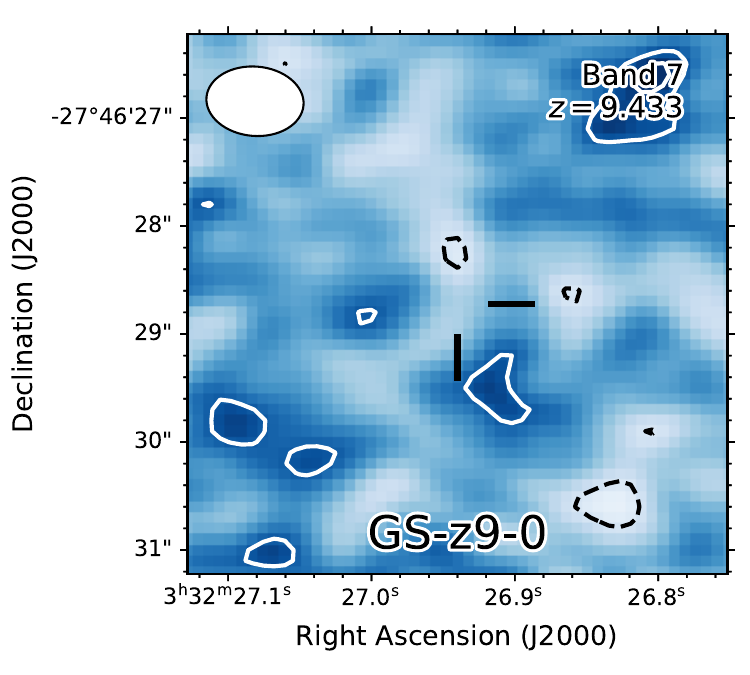}
\includegraphics[width=0.245\linewidth]{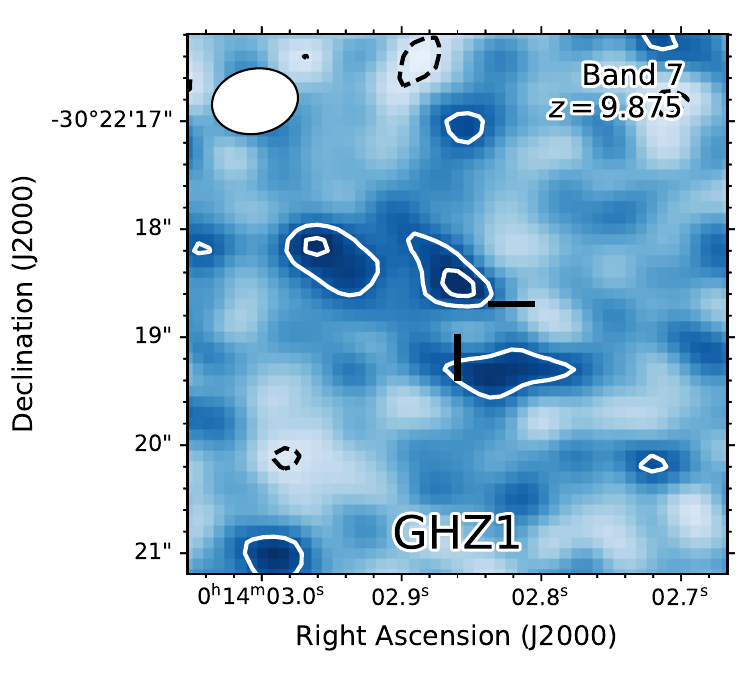}
\includegraphics[width=0.245\linewidth]{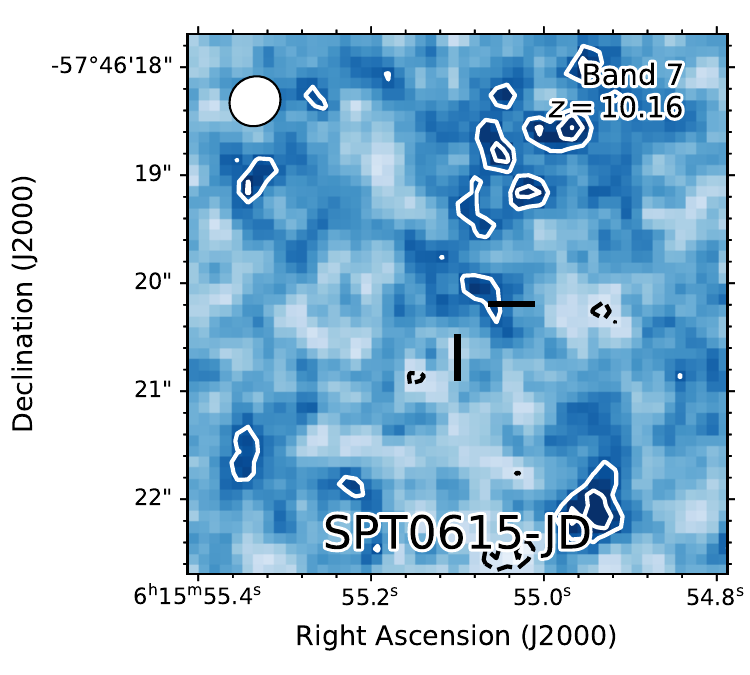}
\includegraphics[width=0.245\linewidth]{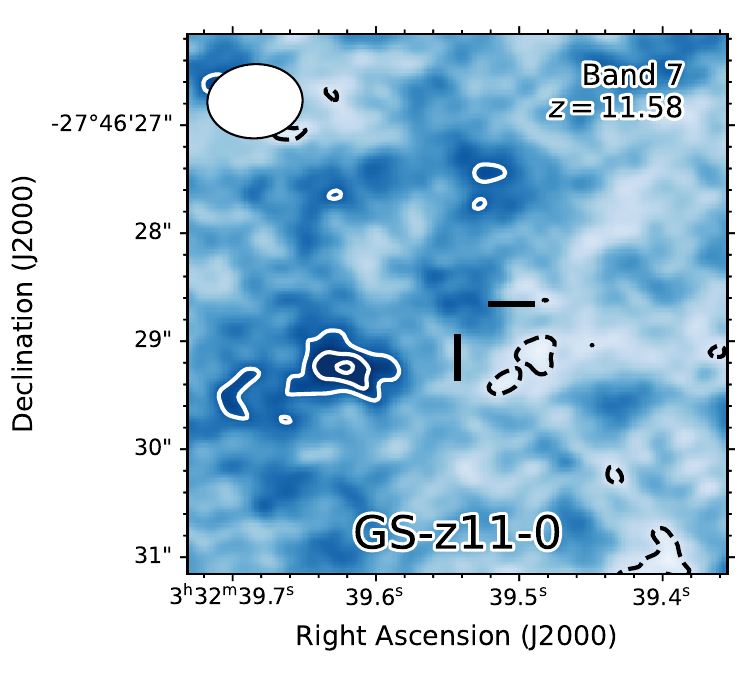}
\includegraphics[width=0.245\linewidth]{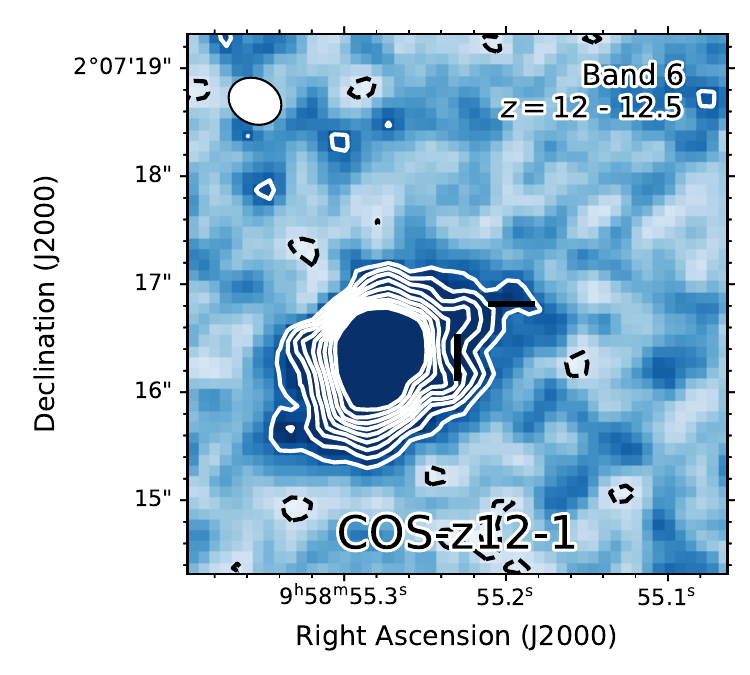}
\includegraphics[width=0.245\linewidth]{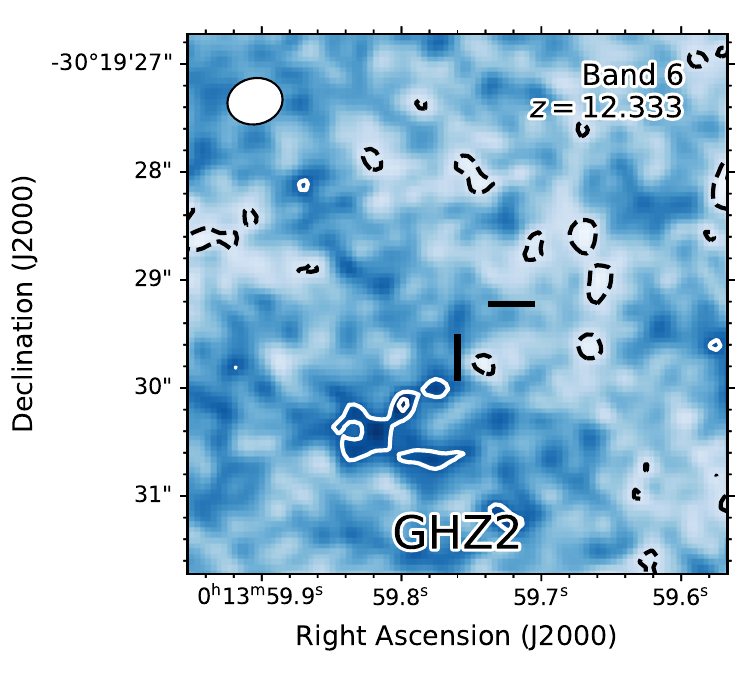}
\includegraphics[width=0.245\linewidth]{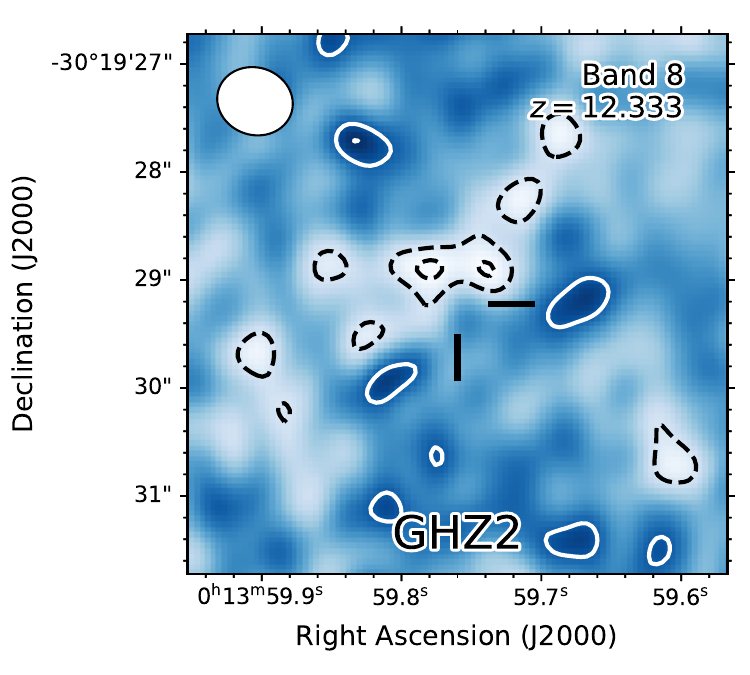}
\includegraphics[width=0.245\linewidth]{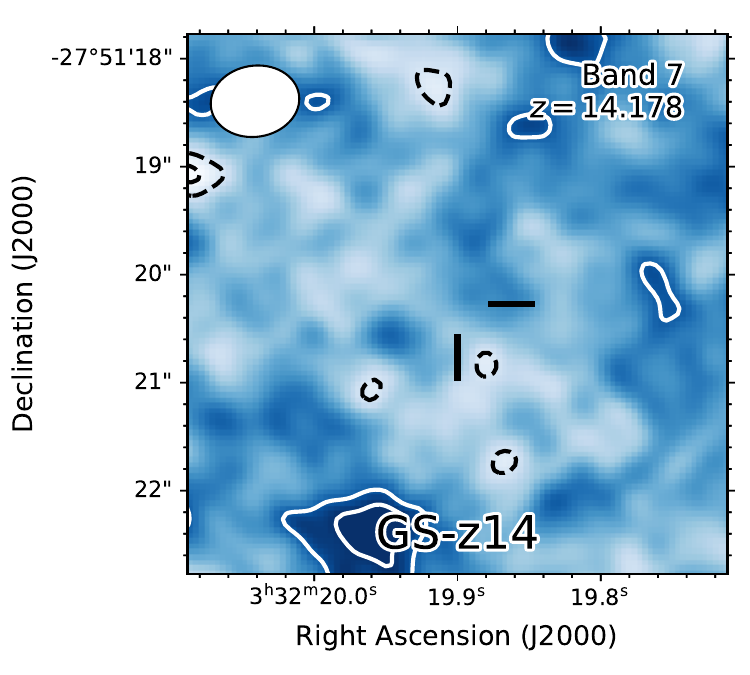}
\includegraphics[width=0.245\linewidth]{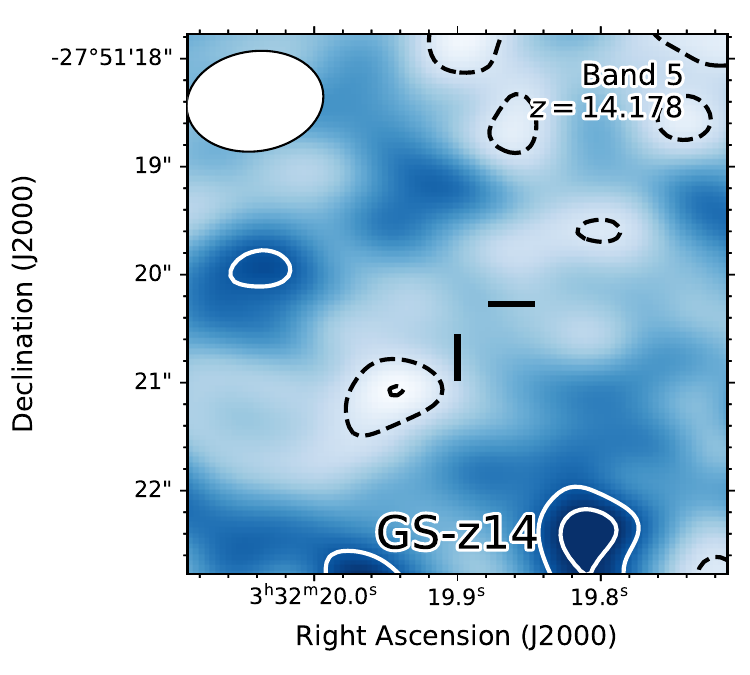}
\includegraphics[height=0.22\linewidth]{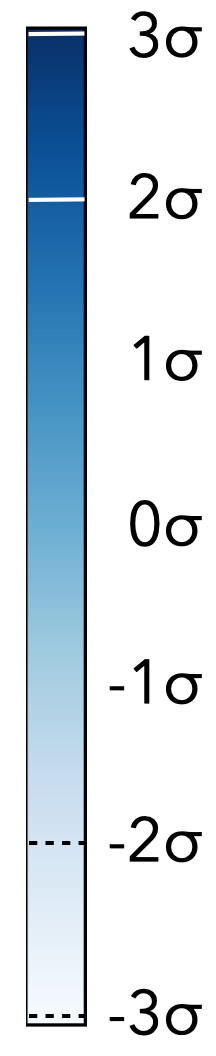}
    \caption{Compilation of all deep dust observations of $z > 8$ sources. Contour levels are drawn at 2, 3, 4, $5 \sigma$..., with dashed contours indicating negative values. The source position is indicated by two off-centre black lines, and the beam is shown in the top-left. }
    \label{fig:continuumImages}
\end{figure*}

\subsubsection{REBELS-24}
This source (also known as UVISTA-Y-005) was observed as a pilot source of the Reionization Era Bright Emission Line Survey (REBELS) ALMA Large Program \citep{Schouws2022}, with additional observations in the full-scale REBELS program \citep{Bouwens2022REBELS,Inami2022}. It was initially identified from very deep optical, near-IR, and Spitzer/IRAC observations obtained across the 2 deg$^2$ COSMOS/UltraVISTA field with data from COSMOS \citep{Scoville2007,Capak2007}, CFHT-LS \citep{Erben2009,Hildebrandt2009}, UltraVISTA \citep{McCracken2012}, SPLASH \citep{Capak2013}, and SMUVS \citep{Caputi2017,Ashby2018}. Even though it had a high photometric redshift accuracy at $z_{\rm phot} \approx 8.6$ (later updated to $z_{\rm phot} \approx 8.35$ in \citealt{Bouwens2022REBELS}) using the unresolved emission of \textsc{[O\,iii]}$\lambda5007$ and $\mathrm{H\beta}$ lines, no far-infrared spectral line was seen with high enough significance to provide a robust spectroscopic redshift. The initial photometry used in the REBELS survey provided a relatively uncertain stellar mass of $\log_{10} M_{\ast} / M_{\odot} = 8.89 \pm 0.68$. Additional NIRSpec-IFU ($R = f/\delta f =100$) \textit{JWST} observations (Cycle 1 GO program \#2659; P.I. John Weaver) have revealed the rest-frame optical emission lines confirming its redshift at $z = 8.21$. 

\subsubsection{GS-z9-3}
% Description of source
GS-z9-3 (also known as UDF-s2-1) has been first identified in the \textit{Hubble} Ultra Deep Field (UDF), and has a robust spectroscopic redshift of $z_{\rm spec} = 8.228$, which was found using NIRCam/grism observations as part of the First Reionization Epoch Spectroscopically Complete Observations (FRESCO; \citealt{Oesch2023Fresco}). Subsequent modeling of its \textit{HST} and \textit{JWST} data \citep{Laporte2023} using \textsc{BAGPIPES} \citep{Carnall2018}, including medium-band filters (Cycle 1 GO program \#1963, P.I. Christina Williams) reveals a massive galaxy ($\log_{10}$ $M_{\ast} /$M$_{\odot}$ = $9.19^{+0.07}_{-0.06}$) with a young stellar component ($30 \pm 10$~Myr) and a high metallicity ($Z = 0.94^{+0.13}_{-0.18}$~Z$_{\odot}$).
% Description of the submm observations
GS-z9-3 has been observed with ALMA as part of the project 2017.1.00486.S (P.I. Richard Ellis) in an effort to detect \oiiil{} emission, originating from the ionized gas surrounding massive O and B-type stars, while also providing a deep constraint on the dust mass. The observations are centered on a rest-frame frequency of 349~GHz based on earlier photometric estimates, but do not cover the true \oiiil{} frequency.

\subsubsection{ID4590}
% Description of source
This source was first identified in the JWST Early Release Observations to have a spectroscopic redshift of $z = 8.496$ \citep{Nakajima2023}. The analysis of the \textit{JWST} observations suggest a metallicity of $12 + \log{\rm (O/H)} = 7.26 \pm 0.18$ \citep{Heintz2023}. 
% Description of NIR properties
% Description of the submm observations
Subsequent ALMA studies by \citet{Fujimoto2024} revealed a tentative \cii{} and \oiii{} detections. Neither band -- at rest-frame wavelengths 160 and 90~$\mu$m -- showed a continuum detection, providing an upper limit on its dust mass of $\log(\mu M_\mathrm{dust} / M_\odot) = 6$.

\subsubsection{MACS1149-JD1}

% Description of source

% Description of NIR properties

% Description of the submm observations
MACS1149-JD1 is a strongly gravitationally lensed galaxy at $z=9.11$ with a spectroscopically confirmed redshift found through its \oiiil{} emission by \citet{Hashimoto2018Natur.557..392H}, first found behind the Hubble Frontier Field cluster MACS1149 \citep{Zheng2012JD1discovery}. Deeper resolved observations of the \oiii{} line were presented by \citet{Tokuoka2022}, who also measured a sensitive upper limit on the dust continuum emission of JD1. Its high lensing magnification is $\mu = 10.5$ of this intrinsically-faint galaxy (${\rm M_{UV} = -19.6}$; \citealt{Bouwens2022REBELS}), and while the \cii{} emission from JD1 was targeted, it was not detected \citep{Laporte2019}, with later analysis indicating a low-significance line feature around the \cii{}-predicted frequency \citep{Carniani2020}. The stellar mass measurement of JD1 is estimated to be $\log(M_\star/M_\odot) = 7.47 \pm 0.05$ \citep{marconcini2024}, and the metallicity is measured to be $12 + \log(\mathrm{O/H}) = 7.82 \pm 0.07$ \citep{morishita2024}.

\subsubsection{GS-z9-0}
% Description of source
JADES-GS-z9-0 (hereafter GS-z9-0, previously known as GS-z10-1), a luminous galaxy spectroscopically confirmed at $z = 9.433$ in the HUDF \citep{Bunker2024}. This source was originally identified as a robust high-redshift galaxy candidate within the CANDELS GOODS-S field on the basis on its red J125 - H160 color by \cite{Oesch2014} (the source was known as GS-z10-1 in that work, and subsequently, on the ALMA archive). Deep ($\sim 130$~hr on-source) \textit{JWST} data with NIRSpec and NIRCam provide exquisite spectroscopy and photometry that enable modeling using \textsc{BEAGLE} and \textsc{BAGPIPES}. The resulting fitting \citep{Curti2024} reveals a stellar mass of $\log_{10}$ $M_{\ast} /M_{\odot}$ = $8.17^{+0.06}_{-0.06}$, star-formation rate of $5 \rm\, M_{\odot}/yr$, low dust attenuation ($\rm A_V = 0.002 - 0.05$) and metallicity of $0.06 \pm 0.06 \rm \, Z_{\odot}$. 
% Description of NIR properties
% Description of the submm observations
As part of a campaign to search for the \oiiil{} emission of GS-z9-0, ALMA observed the $\sim 310$~GHz regime using Band 7 (project code 2022.1.01401.S; P.I. S. Serjeant) in two tunings aiming to target the \oiii{} emission line. However, the ALMA observations did not cover the exact \oiii{} frequency, but do provide a deep upper limit on the dust mass. 

\subsubsection{GHZ-1}
% Description of source
GHZ-1 (also known as GLASS-z10, and optimistically GLASS-z11 in earlier reports) is one of the first $z > 10$ galaxy candidates revealed by the \textit{JWST} as part of the GLASS \textit{JWST} Early Release Science observations (program \#1324; P.I. Tommaso Treu). Analysis of these \textit{JWST} observations swiftly revealed two high-$z$ galaxies \citep{Naidu2022,Castellano2022}, which were subsequently studied with ALMA observations to identify their spectroscopic redshift through \oiiil{} emission. Neither \oiii{} nor dust was detected \citep{Yoon2023}, but subsequent NIRSpec observations of high-redshift sources in the GLASS fields identified its redshift to be $z = 9.875$ \citep{Napolitano2025}. This redshift indicated that the \oiiil{} was not covered by the spectral set-up of the ALMA observations.

% Description of NIR properties

% Description of the submm observations

\subsubsection{GN-z11}
GN-z11 is a UV-luminous Lyman-break galaxy found in the CANDELS GOODS/North field \citep{Oesch2014}. Initial \textit{HST} GRISM observations confirmed the redshift at $z = 11.1$ \citep{Oesch2016}, with subsequent \textit{JWST} imaging revealing the true redshift to lie at $z=10.60$ \citep{Bunker2023}. GN-z11 has a stellar mass of $\log_{10} M_{\ast} / M_{\odot} = 8.73 \pm 0.06$, and a metallicity of $12\pm2$~$Z_{\odot}$ \citep{Bunker2023}. Its high declination ($> 60^{\circ}$) required the use of the NOEMA to target its \cii{}, with some observations targeting the earlier redshift solution. Through a combined seven observing sessions, its \cii{} and underlying dust continuum emission were targeted \citep{Fudamoto2024}. Neither were detected, but a deep upper limit on this UV-bright galaxy is provided. 

% https://arxiv.org/pdf/2309.02493 https://ui.adsabs.harvard.edu/link_gateway/2023A%26A...677A..88B/PUB_PDF
% Description of source

% Description of NIR properties

% Description of the submm observations

\subsubsection{GS-z11-0}
GS-z11-0 (also known as UDFj-39546284; \citealt{2011Natur.469..504B, 2013ApJ...763L...7E, 2013ApJS..209....3K}) was discovered in the Hubble Ultra Deep Field \citep[HUDF;][]{2006AJ....132.1729B}. Following its photometric characterisation using NIRCam in the JWST Advanced Deep Extragalactic Survey \citep[JADES;][]{2023arXiv230602465E} as a promising $z > 10$ candidate, it was selected for deep spectroscopic follow-up with NIRSpec. This spectroscopic analysis revealed the Lyman-break feature with high fidelity, confirming it as one of the four first spectroscopically confirmed $z > 10$ galaxies \citep{2023NatAs...7..622C, 2023NatAs...7..611R, Hainline2024}.
In Cycle 10, band-6 and -7 observations of GS-z11-0 were taken as part of programme 2023.1.00336.S (PIs: Joris Witstok \& Renske Smit) primarily to target the \oiiil{} line \citep{Witstok2025}.

% Description of NIR properties

% Description of the submm observations

\subsubsection{GHZ-2}
GHZ2 (also known as GLASS-z12, and optimistically GLASS-z13 in earlier reports) is the highest-redshift candidate found in the initial study of the ERS GLASS program (program \#1324; P.I. Tommaso Treu). 
Similar to GHZ-1, this source was swiftly suspected to be a high-$z$ galaxy \citep{Naidu2022,Castellano2022}, and was subsequently studied with ALMA observations to identify their spectroscopic redshift through \oiiil{} emission \citep{Bakx2023,Popping2023}. Initially, a low-significance tentative \oiiil{} feature was reported, but subsequent NIRSpec \citep{Castellano2024} and MIRI \citep{Zavala2025MIRI} observations revealed the spectroscopic redshift of $z = 12.333$ to coincide with a lower-significance $3 \sigma$ \oiiil{} feature. Its stellar mass ($\log M_{\ast}/M_{\odot} = $~9.05$_{-0.25}^{+0.10}$) is reported in \citet{Castellano2024} and studies of its metallicity ($Z / Z_{\odot} = 5$--11~per cent, depending on models) in \citet{Calabro2024}. Additional spectroscopically-targeted observations of this feature revealed $5 \sigma$ \oiiil{} emission, as well as an upper limit on the \oiii{} 52$\mu$m emission line \citep{Zavala2024OIII,Zavala2025MIRI}, and accompanying deep upper limits on the dust emission at 90 and 50~$\mu$m \citep{Mitsuhashi2025}. This data is combined with the data that aims to detect the \niii{}~57~$\mu$m emission (2024.1.01645; P.I. Renske Smit). The expected \niii{}-emitting region is masked ($\pm 500$~km/s; Smit et al. in prep.), and combined with the continuum surrounding the \oiii{}~52~$\mu$m emission for a deep limit at short wavelengths.

% Description of source

% Description of NIR properties

% Description of the submm observations

\subsubsection{GS-z14-0 }
JADES-GS-z14-0 (GS-z14-0) is currently the most distant galaxy with a confirmed spectroscopic redshift at $z=14.18$. Following its spectroscopic confirmation with NIRSpec \citep{Carniani2024highzSources}, it was detected in \oiiil{} emission with ALMA \citep{Carniani2024EventfulLife,Schouws2024OIII}. Even though GS-z14-0 is relatively massive at $\log(M_\star/M_\odot) = 8.78_{-0.10}^{+0.09}$ \citep{Carniani2024highzSources,Helton2024}, no dust continuum is seen at 90~$\mu$m. A potential explanation of this could be dusty outflows \citep{Ferrara2024EventfulLife,Ferrara2025AFM}. Deeper follow-up observations targeting \ciil{} provide a deep upper limit while also not detecting the dust emission \citep{Schouws2025}.

\subsection{Sources that are not stacked}

\subsubsection{MACS0416-Y1}
% Description of source
% Description of NIR properties
MACS0416\_Y1 is a Lyman-Break Galaxy (LBG) originally identified as one of the most distant galaxies behind the galaxy cluster MACSJ0416.1$-$2403, which is one of the Hubble Frontier Fields, with a magnification of $\mu = 1.5$ \citep{Kawamata2016, Rihtar2024arXiv240610332R} based on improved estimates from the \textit{JWST}. Subsequent ALMA observations revealed the spectroscopic redshift of MACS0416\_Y1 to be $z = 8.312$ through its \oiiil{} emission \citep{Tamura2019}. Importantly, these observations detected bright dust emission $137 \pm 26$~$\mu$Jy surrounding the \oiii{} line, and make it the furthest object with detected dust emission.

The lack of dust emission at 160~\micron{} hinted at a warm dust temperature in excess of 80~K \citep{Bakx2020CII,Algera2024}, while the moderately-resolved \cii{} emission suggests a rotationally-stable star-forming system. Higher-resolution follow-up \citep{Tamura2023} in \oiiil{} emission resolved clumpy and non-rotating \hii{} regions within MACS0416\_Y1 in three distinct parts, where the dust and UV-emission appears to originate from separate locations, explaining the relatively-blue UV emission seen in MACS0416\_Y1 ($\beta_{\rm UV} \approx -1.8$ where F$_{\lambda} \propto \lambda^{\beta_{\rm UV}}$). The high-resolution morphology from the \textit{JWST} provides early indications that this source is a merging system \citep{Ma2024ApJ...975...87M}, and spectroscopic analysis of the \textit{JWST} data \citep{Harshan2024} further suggests it contains a chemically-evolved interstellar medium reminiscent of $z = 2$ star-forming galaxies \citep{Sanders2020MOSDEF}. Meanwhile, deep radio observations with the Very Large Array \citep{Jones2024} fail to find molecular gas through carbon-monoxide lines. 

% Description of the submm observations

\subsubsection{SPT-0615-JD}
% Description of source
SPT0615-JD (also known as Cosmic Gems Arc) was discovered by \cite{Salmon2018} in the Reionization Lensing Cluster Survey (RELICS) Hubble Treasury program \citep{Coe2019}. It lies behind the galaxy cluster SPT-CL J0615-5746, and is magnified by roughly $\mu = 120$ with a lensing arc of $\sim 5$ arcseconds. Although SPT0615-JD initially did not have a spectroscopic redshift yet, \textit{JWST} imaging agrees with a high-redshift nature of $z_{\rm phot} = 10.2 \pm 0.2$ \citep{Bradley2024,Hsiao2024} with an estimated stellar mass of $\log_{10} M_{\ast}/$M$_{\odot} = 7.47$ $\pm$ 0.18. {\color{referee2} Subsequent NIRSpec spectroscopy has since confirmed its redshift to be 9.625 \citep{Messa2025}.} Previous attempts with ALMA to identify its spectroscopic redshift through \oiiil{} have furthermore provided deep sub-mm imaging (projects 2018.1.00295.S and 2019.1.00327.S; P.I. Yoichi Tamura). 
Since the optical emission of this source is extended well beyond the ALMA beam, even in heavily tapered ALMA imaging, we do not include this source in our stack, but document its existence (as already done in \citealt{Bradley2024}) for posterity and completeness, as it is a high-redshift galaxy with deep ALMA imaging.

% Description of NIR properties

% Description of the submm observations

\subsubsection{COS-z-0/COS-z12-1}
% Description of source
COS-z-0 (also known as COS-z12-1) is located in the COSMOS field and was detected as part of the COSMOS-Web study \citep{Casey2024}. COSMOS-Web is one of the largest-area surveys with JWST, making it likely one of the brightest galaxies in the $z > 10$ Universe. As one of the brightest, highest-redshift candidates, it enjoyed extensive ALMA follow-up to target the \oiiil{} emission line (2023.A.00003.S; P.I. Caitlin Casey). The deep \textit{JWST}-available photometry provide confidence in its high-redshift nature, and identified its stellar mass to be $\log_{10} M_{\ast} / M_{\odot} = 9.60 \pm 0.11$. However, a nearby ($\sim 1$~arcsec) dust-emitting galaxy complicates the study of its direct dust emission, even after attempted subtraction of the foreground galaxy. Consequently, we exclude this source from this stacking study.

\subsection{High-redshift unconfirmed targets}
Several tentative high-redshift ($z > 8$) sources have enjoyed deep ALMA follow-up, but are suspect of being low-redshift interlopers. For completeness, we list the existing data below, but do not include them in our stacking.

\subsubsection{COSMOS\_20646}
This target is identified as a bright high-redshift galaxy found in a CANDELS field \citep{Finkelstein2022CANDELS,Tacchella2022}. It has a very high estimated stellar mass ($\log_{10} M_{\ast}/$M$_{\odot} = 9.77 \pm 0.19$; \citealt{Tacchella2022}), and has several features in its optical and near-infrared photometry that leave a 4~per cent chance of a lower-redshift interloper ($z \sim 2.5$). In an effort to confirm its high-redshift nature, ALMA observations targeted the expected region where the \oiiil{} emission was located in project 2019.1.00397.S (P.I. Takuya Hashimoto; Arai et al. in prep.), although no line emission is apparent. As a potential low-redshift interloper, this source is excluded in this stacking study.

\subsubsection{UDS-18697}
Similar to COSMOS\_20646, UDS-18697 is identified in the CANDELS field with an estimated high stellar mass of $\log_{10} M_{\ast}/M_{\odot} = 11^{+0.4}_{-0.2}$ \citep{Finkelstein2022CANDELS,Tacchella2022}. Deep ALMA observations (2019.1.00397.S; P.I. Takuya Hashimoto and 2022.1.01562.S; P.I. Seiji Fujimoto; Arai et al. in prep.) provide deep upper limits but no confirmation of the high redshift. Subsequent \textit{JWST} observations (\#1758 P.I Finkelstein) used NIRSpec to target this bright galaxy. A preliminary investigation of the deep NIRSpec data do not reveal any bright emission lines characteristic of the high-redshift solution. As a potential low-redshift interloper, this source is excluded from this study.

\subsubsection{COSMOS2020\_441697 and COSMOS2020\_1356755}
These galaxies are identified as high-redshift candidates in the COSMOS field \citep{Kauffmann2022}. These sources were observed with the same Cycle 1 GO program as REBELS-24 (\#2659; P.I. John Weaver), but lack the immediately-apparent emission lines that provide a high-redshift verification. Even though both fields enjoy deep ALMA observations in an effort to confirm their high-redshift nature through the \oiiil{} and dust emission (2022.1.01562.S; P.I. S. Fujimoto and 2021.1.00389.S; Takuya Hashimoto; Arai et al. in prep.), the lack of verification of their high-redshift nature led us to exclude these sources from the stacking study.

\subsubsection{2140+0241-37}
2140+0241-37 (also known as par2139+0241~1709) is identified through the Brightest of the Reionization Galaxies (BoRG) \textit{HST} survey \citep{Calvi2016}, which is the combined multi-year effort of pure-parallel near-IR and
optical imaging with the Wide Field Camera 3. The photometric redshift places this source at $z_{\rm phot} = 10.5$, providing strong incentive to search for its \oiiil{} emission using ALMA (2019.1.00397.S; P.I. Takuya Hashimoto; Arai et al. in prep.). Subsequent analysis in \cite{Rojas-Ruiz2020}, where the source is named par2139+0241~1709, re-analysed the Spitzer/IRAC emission and found it not a likely high-redshift candidate, as it has a turnover in the spectrum at an observed wavelength of $\sim 3\, \mu$m. Together with the complex morphology of the source at longer wavelengths, this is consistent with a $z < 2$ stellar emission spectrum. 

\subsubsection{HD1}
HD1 is a high-redshift candidate ($z_{\rm phot} = 13.3$) found through the 2.3~sqr. deg. survey with deep Subaru data, together with UltraVISTA, COSMOS, and SXDS data \citep[][and ref. therein]{Harikane2022}. Subsequent ALMA imaging (2019.A.00015.S; P.I. Akio Inoue) provided a tentative $4 \sigma$ \oiiil{} at the expected redshift, but subsequent deep \textit{JWST} imaging revealed this source to be a low-redshift interloper instead \citep{Harikane2025}. 

% Description of source

% Description of NIR properties

% Description of the submm observations

\section{Stack in sub-mm flux}
\label{sec:stacksubmmflux}
\begin{figure*}
\begin{minipage}{.4\textwidth}
  \includegraphics[width=1\linewidth]{graphs/stack_reprojectCont_all.fits.pdf}
\end{minipage}%
\begin{minipage}{.6\textwidth}
\includegraphics[width=0.33\linewidth]{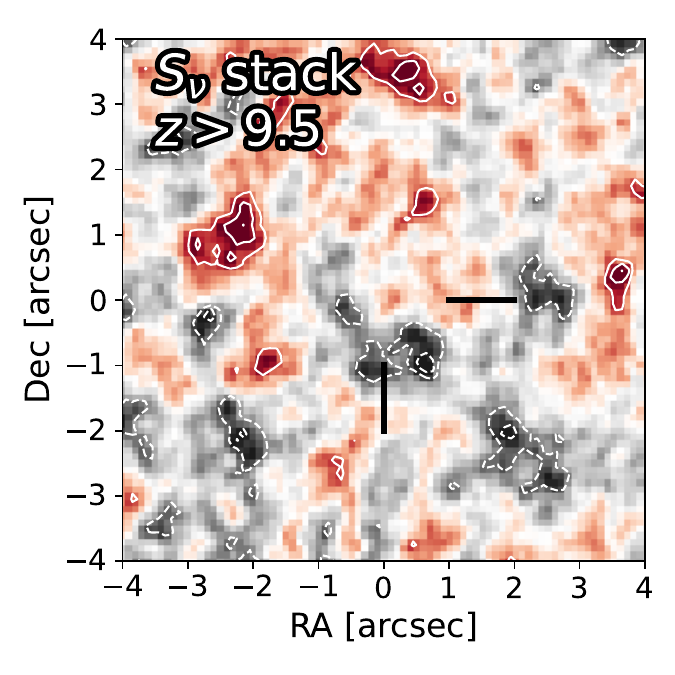}
\includegraphics[width=0.33\linewidth]{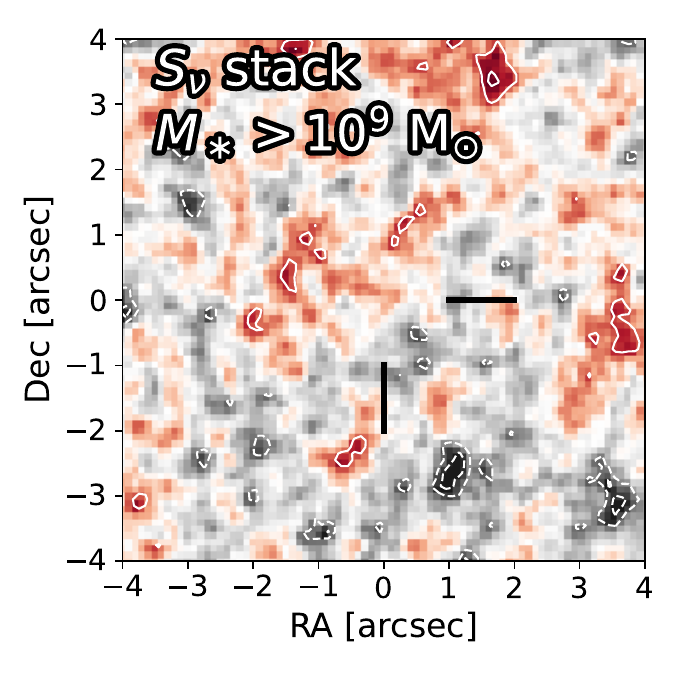}
\includegraphics[width=0.33\linewidth]{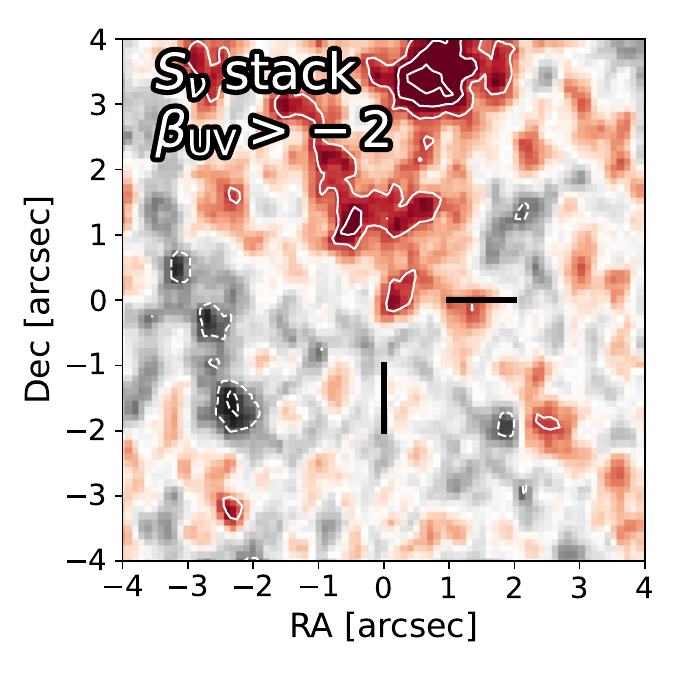}\\
\includegraphics[width=0.33\linewidth]{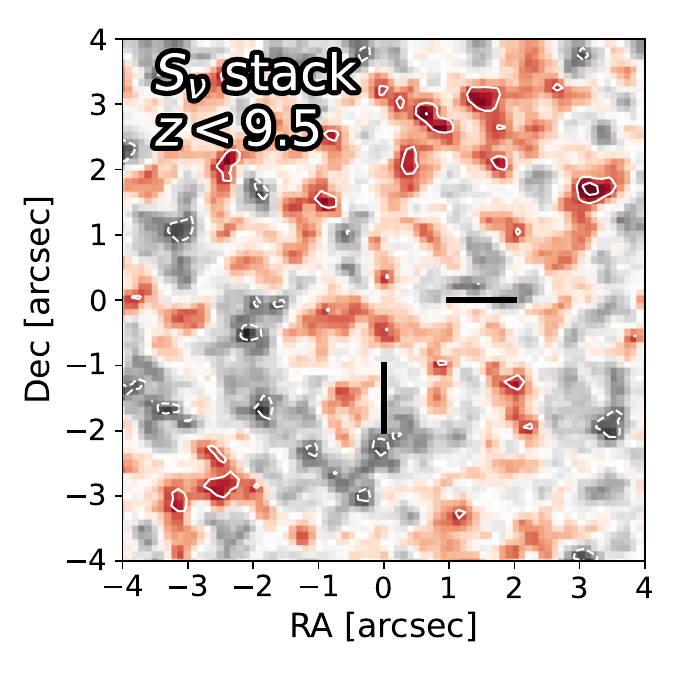}
\includegraphics[width=0.33\linewidth]{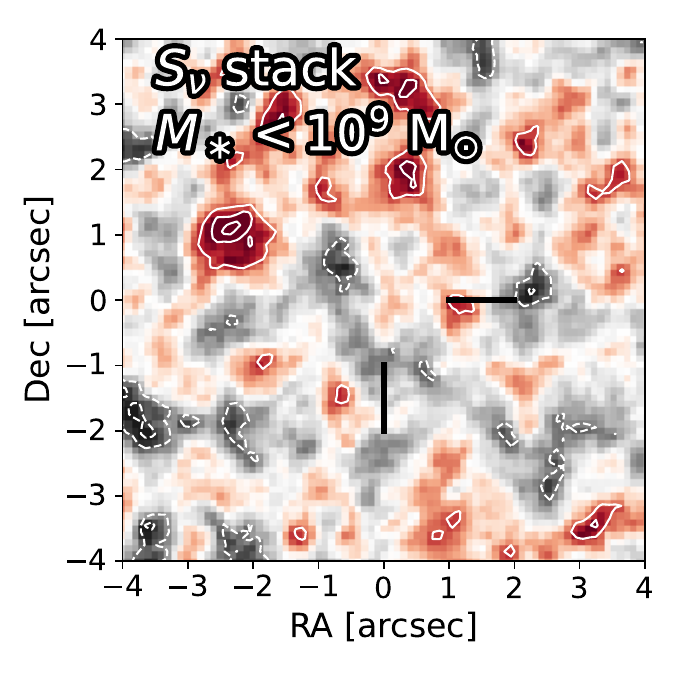}
\includegraphics[width=0.33\linewidth]{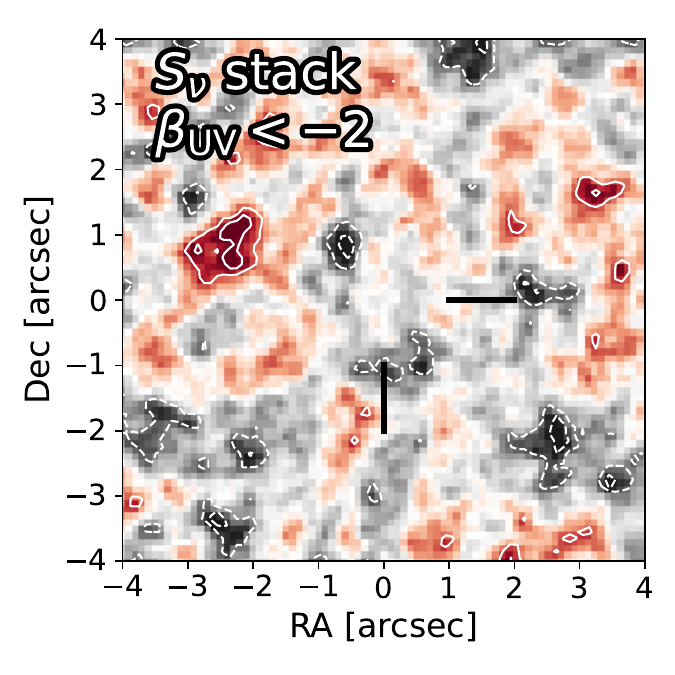}
\end{minipage}
\caption{The full flux density stack of all data (left), and the sub-sets (left-to-right) are in redshift, stellar mass, and $\beta_{\rm UV}$ bins, where the top row contains the higher, and the bottom row contains the lower quantities. }
    \label{fig:SnuMassStacks}
\end{figure*}

Figure~\ref{fig:SnuMassStacks} shows the stacking results weighted by flux density in units of Jy / beam. 

\section{Stack in dust mass}
\begin{figure*}
\begin{minipage}{.4\textwidth}
  \includegraphics[width=1\linewidth]{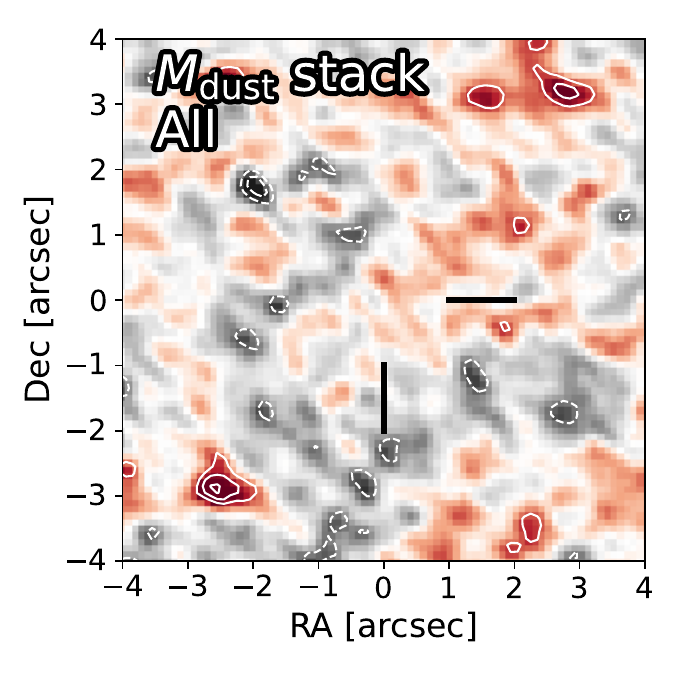}
\end{minipage}%
\begin{minipage}{.6\textwidth}
\includegraphics[width=0.33\linewidth]{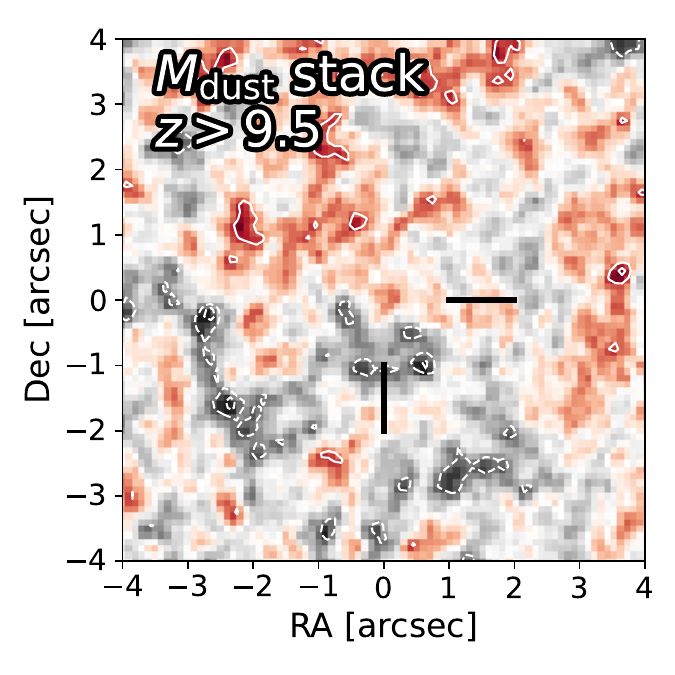}
\includegraphics[width=0.33\linewidth]{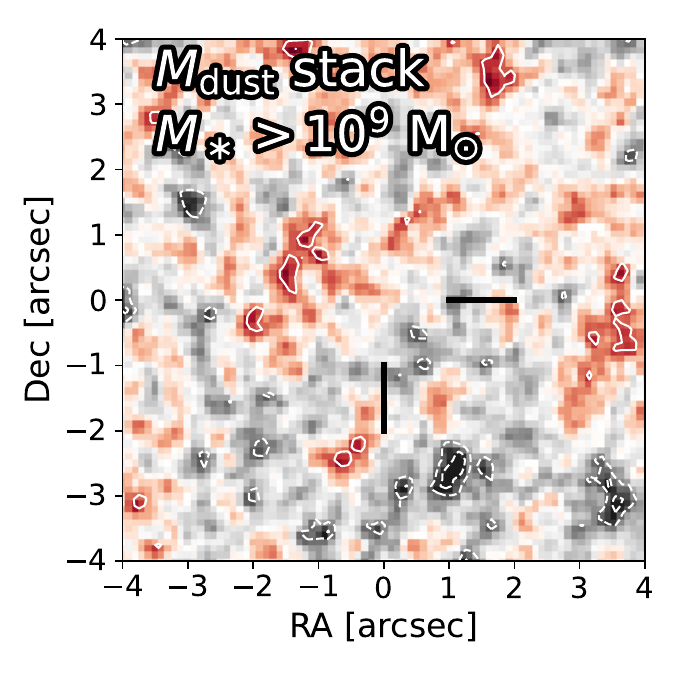}
\includegraphics[width=0.33\linewidth]{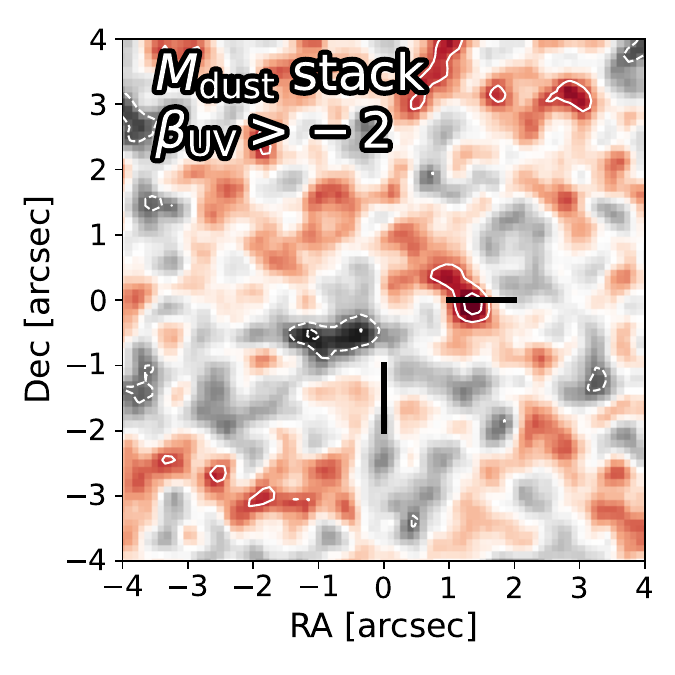}\\
\includegraphics[width=0.33\linewidth]{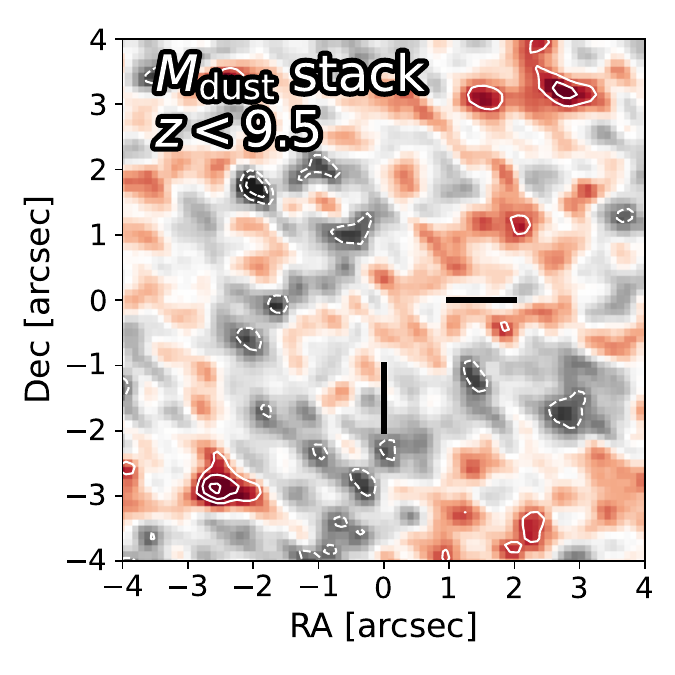}
\includegraphics[width=0.33\linewidth]{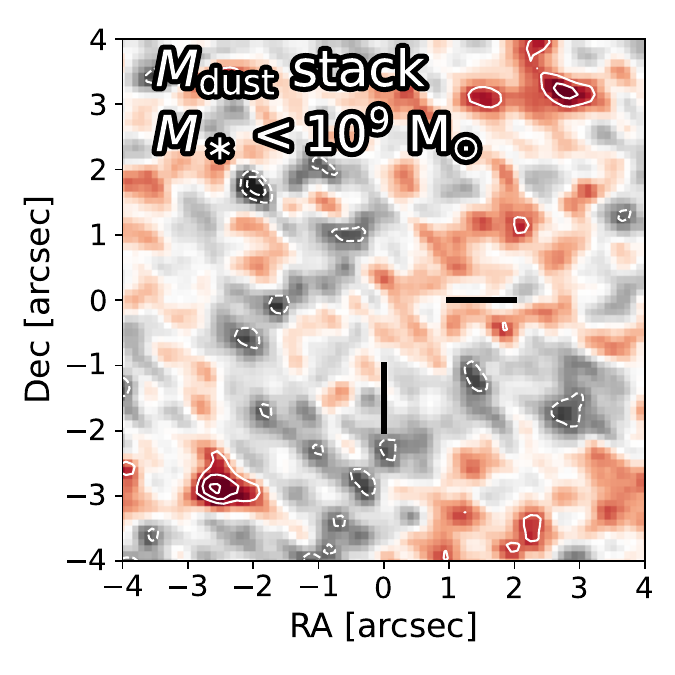}
\includegraphics[width=0.33\linewidth]{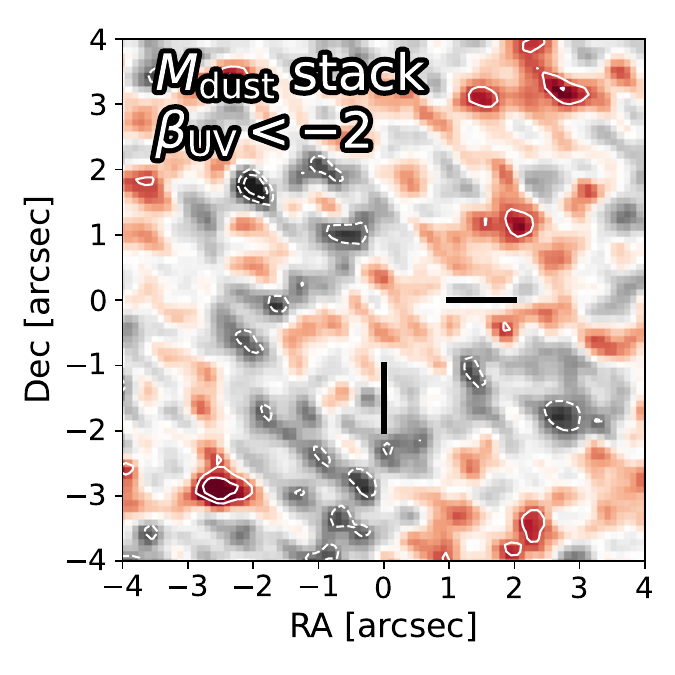}
\end{minipage}
\caption{The full dust mass stack of all data (left), and the sub-sets (left-to-right) are in redshift, stellar mass, and $\beta_{\rm UV}$ bins, where the top row contains the higher, and the bottom row contains the lower quantities. }
    \label{fig:DustMassStacks}
\end{figure*}

Figure~\ref{fig:DustMassStacks} shows the stacking results weighted by dust mass in units of 1 / beam. 

\section{Stack in dust-to-stellar mass ratio}
\label{sec:stackdusttostellarmass}
\begin{figure*}
\begin{minipage}{.4\textwidth}
  \includegraphics[width=1\linewidth]{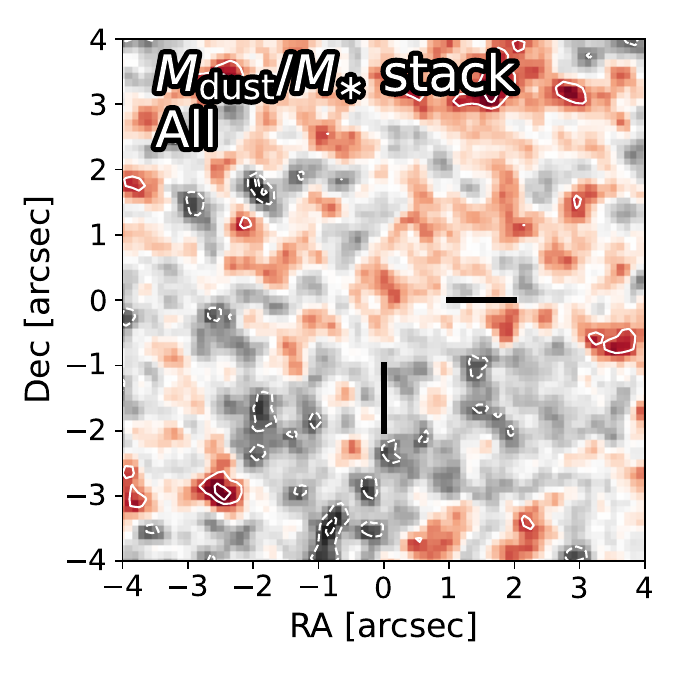}
\end{minipage}%
\begin{minipage}{.6\textwidth}
\includegraphics[width=0.33\linewidth]{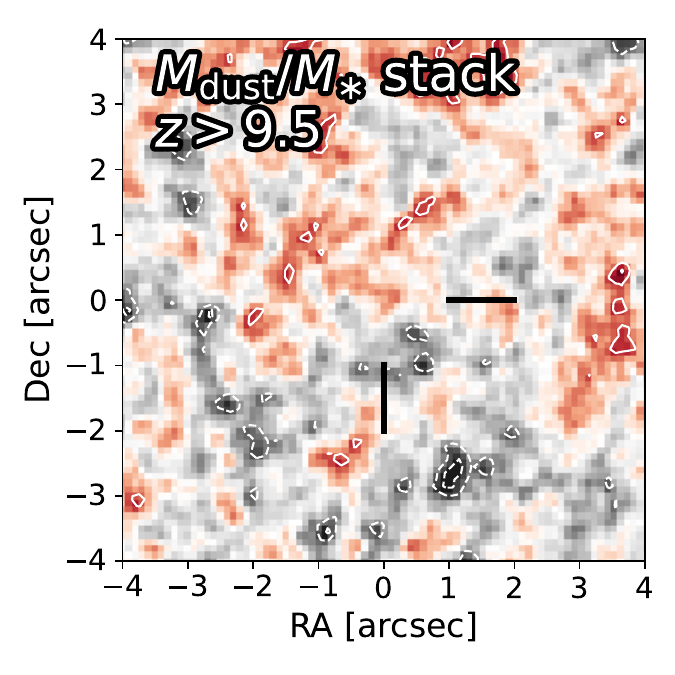}
\includegraphics[width=0.33\linewidth]{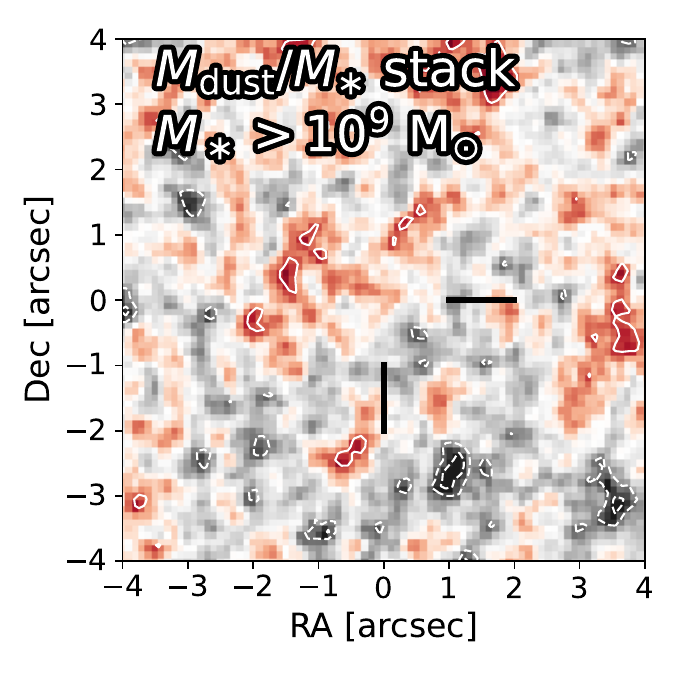}
\includegraphics[width=0.33\linewidth]{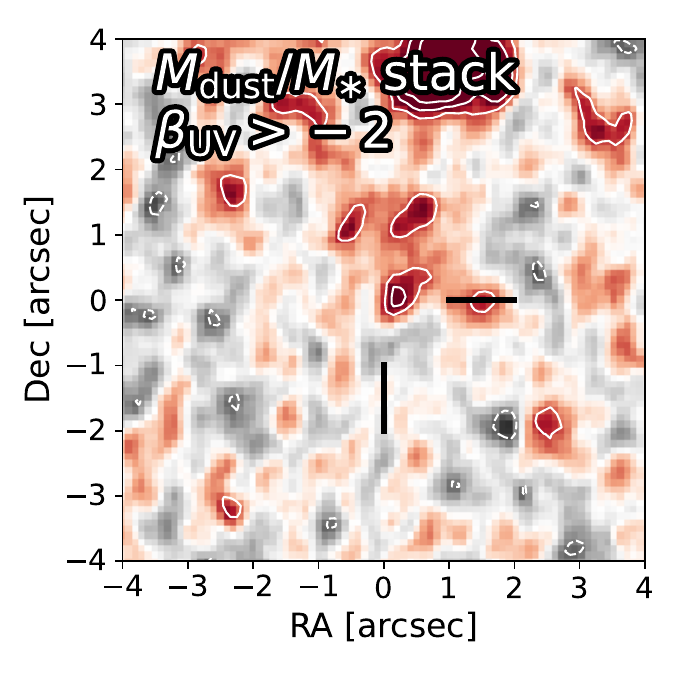}\\
\includegraphics[width=0.33\linewidth]{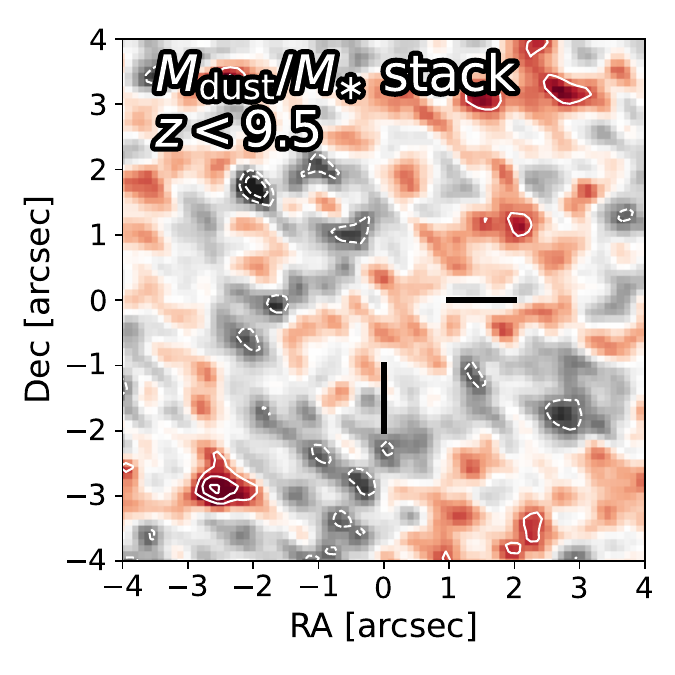}
\includegraphics[width=0.33\linewidth]{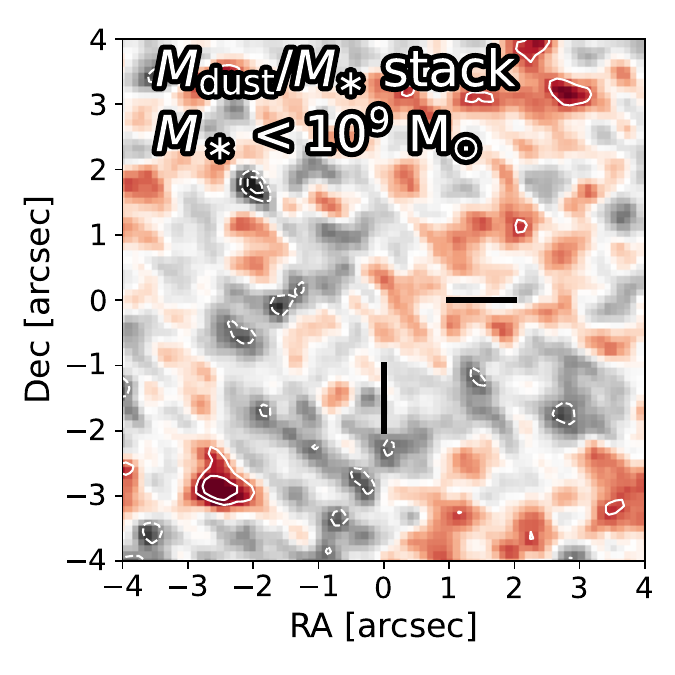}
\includegraphics[width=0.33\linewidth]{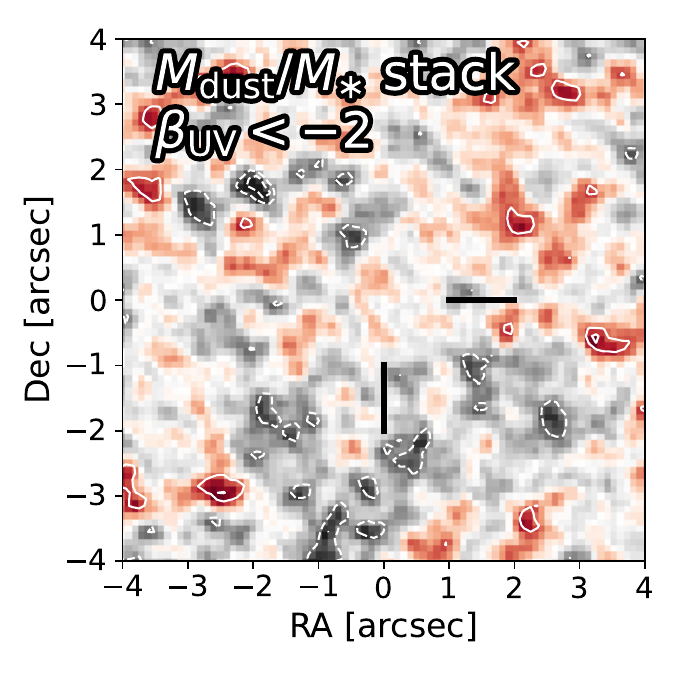}
\end{minipage}
\caption{The full dust-to-stellar mass ratio stack of all data (left), and the sub-sets (left-to-right) are in redshift, stellar mass, and $\beta_{\rm UV}$ bins, where the top row contains the higher, and the bottom row contains the lower quantities. }
    \label{fig:DustToStarMassStacks}
\end{figure*}

Figure~\ref{fig:DustToStarMassStacks} shows the stacking results weighted by dust-to-stellar mass ratio in units of 1 / beam. 

{\color{referee2}
\section{Comparison of dust stacks to models}
Appendix Figure~\ref{fig:Mstar_vs_MstarMdust_stack} shows the dust mass of our stack against reference stacks at cosmic noon \citep{Bouwens2016,Bouwens2020,Dunlop2017,McLure2018,Shivaei2022} and dawn \citep{Algera2023,Bowler2024}. Similarly, the $z = 6 - 8$ dust mass scaling relations detailed in Figure~\ref{fig:MstarMdust} are also shown. For all cosmic noon studies where dust masses were not explicitly given, we assume a 35~K dust temperature and a dust-emissivity index $\beta_{\rm dust}$ of $2$. Even though some studies at cosmic noon only find upper limits after combining hundreds or even thousands of galaxies, these values are in line with the dust scaling relations at $z = 6 - 8$, similar to the $z \sim 2$ solar- and lower-metallicity galaxies studied in \cite{Shivaei2022}. Even though the spectro-photometric catalogues available for stacks at cosmic noon are much larger, the deep limits on the dust mass from our $z > 8$ study still stand out relative to the individual and data at $z < 8$, as well as the dust scaling relations, due to the combined effect of K-correction and the likely higher dust temperatures at $z > 5$. Note that a direct comparison of cosmic noon galaxies (and their stacks) to the $z = 6 - 8$ scaling relations provides additional uncertainties, since these galaxies likely have different metallicities, dust composition and available production/destruction pathways \citep{Boquien2022,Witstok2023,Markov2023,Sommovigo2025}.

\begin{figure*}
    \centering
    \includegraphics[width=0.475\linewidth]{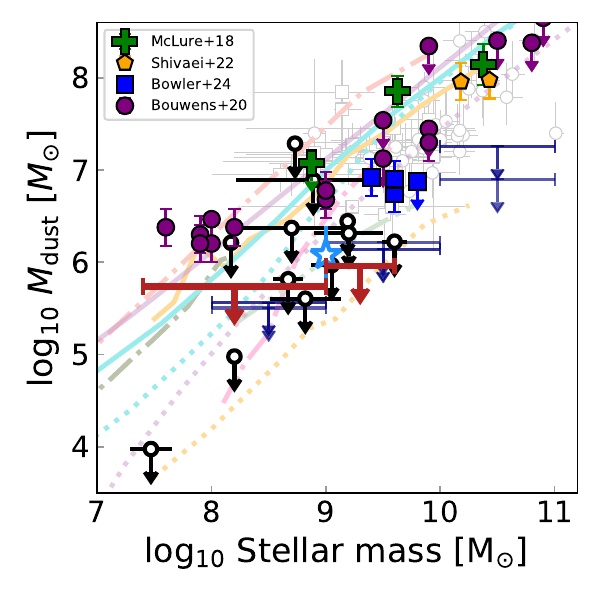}
    \includegraphics[width=0.515\linewidth]{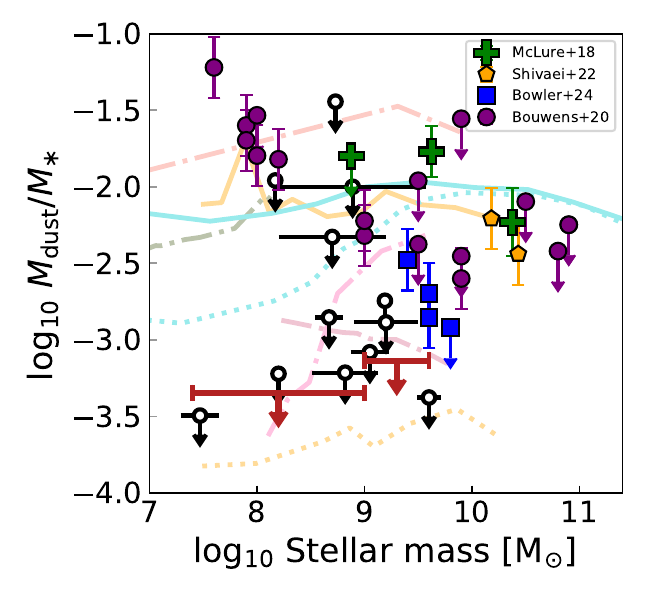}
    \caption{\color{referee2} Similar to Figure~\ref{fig:MstarMdust}, instead compared to previous stacking studies at cosmic noon and dawn. The dust mass and dust-to-stellar mass ratio of the individual galaxies (black circles) and of the stack at two different stellar mass bins ($\log_{10} M_{\ast}/{\rm M_{\odot}} \lessgtr 9$; red upper limits) as a function of stellar mass in the left and right panel, respectively. The upper limits are drawn at $3 \sigma$ and the dust and stellar masses are corrected for lensing. 
    The dust measurements from stacks at cosmic noon  \citep{Bouwens2016,Bouwens2020,Dunlop2017,McLure2018,Shivaei2022} are shown in filled pluses, pentagons and circles, while stacked mass estimates at cosmic dawn are shown in filled squares \citep{Algera2023,Bowler2024}. Note that the left-hand data point of the \citet{Shivaei2022} reflects the lower-metallicity ($\sim 0.5 Z_{\odot}$ galaxies, while the right-hand side data point represents the solar-metallicity galaxies in their sample. 
    These results are compared against individually-detected galaxies at $4.4 < z < 6$ indicated with grey circles \citep{Sommovigo2022} and $6 < z < 8$ galaxies indicated with grey squares \citep{Bakx2021,Sommovigo2022REBELS,Witstok2022,Fudamoto2022DustTemperatures,Algera2024,Algera2024REBELS25,Valentino2024}. The sole source with a dust dust detection at $z = 8.3$ is shown as a blue star \citep{Tamura2019,Tamura2023,Bakx2020CII,Bakx2025}. Dust estimates from stacking experiments are shown as thin blue upper limits \citep{Ciesla2024}, and scaling relations from semi-analytical and hydrodynamical models that account for dust production are shown as trend lines between $z = 7$ to $9.5$ \citep{Popping2017Dustproduction,Imara2018,Vijayan2019,DiCesare2023,Esmerian2024,Triani2021,Dayal2022}. For two dust production models \citep{Popping2017Dustproduction,Vijayan2019}, we show the maximum and average dust production scenarios in solid and dash-dotted lines respectively.
    }
    \label{fig:Mstar_vs_MstarMdust_stack}
\end{figure*}
}

%%%%%%%%%%%%%%%%%%%%%%%%%%%%%%%%%%%%%%%%%%%%%%%%%%

% Don't change these lines
\bsp	% typesetting comment
\label{lastpage}
\end{document}